\documentclass[prc,superscriptaddress,unsortedaddress,twocolumn,showpacs,preprintnumbers,amsmath,amssymb,floatfix]{revtex4}
\usepackage[dvips]{graphicx}
\usepackage{amsmath}
\usepackage{amssymb}
\usepackage{mathrsfs}
\usepackage{multirow}
\usepackage{bm}
\usepackage{color}
\usepackage{threeparttable}

\def\la{\mathrel{\mathpalette\fun <}}

\def\fun#1#2{\lower3.6pt\vbox{\baselineskip0pt\lineskip.9pt
\ialign{$\mathsurround=0pt#1\hfil##\hfil$\crcr#2\crcr\sim\crcr}}}

\newcommand{\beq}{\begin{equation}}
\newcommand{\eeq}{\end{equation}}
\newcommand{\bea}{\begin{eqnarray}}
\newcommand{\eea}{\end{eqnarray}}

\newcommand{\bfi}[1]{\mbox{\boldmath $#1$}}

\newcommand{\vK}{{\bfi K}}

\newcommand{\vs}{{\bfi s}}

\newcommand{\vrr}{{\bfi r}}
\newcommand{\vR}{{\bfi R}}

\begin{document}


\title{Ground-state properties of neutron-rich Mg isotopes}

\author{S. Watanabe}
\email[]{s-watanabe@phys.kyushu-u.ac.jp}
\affiliation{Department of Physics, Kyushu University, Fukuoka 812-8581, Japan}
\author{K. Minomo}
\affiliation{Department of Physics, Kyushu University, Fukuoka 812-8581, Japan}
\author{M. Shimada}
\affiliation{Department of Physics, Kyushu University, Fukuoka 812-8581, Japan}
\author{S. Tagami}
\affiliation{Department of Physics, Kyushu University, Fukuoka 812-8581, Japan}
\author{M. Kimura}
\affiliation{Department of Physics, Hokkaido University, Sapporo 060-0810, Japan}
\author{M. Takechi}
\affiliation{Gesellschaft f\"{u}r Schwerionenforschung GSI, 64291 Darmstadt, Germany}
\affiliation{RIKEN, Nishina Center, Wako, Saitama 351-0106, Japan}
\author{M. Fukuda}
\affiliation{Department of Physics, Osaka University, Osaka 560-0043, Japan}
\author{D. Nishimura}
\affiliation{Department of Physics, Tokyo University of Science, Chiba 278-8510, Japan}
\author{T. Suzuki}
\affiliation{Department of Physics, Saitama University, Saitama 338-8570, Japan}
\author{T. Matsumoto}
\affiliation{Department of Physics, Kyushu University, Fukuoka 812-8581, Japan}
\author{Y. R. Shimizu}
\affiliation{Department of Physics, Kyushu University, Fukuoka 812-8581, Japan}
\author{M. Yahiro}
\affiliation{Department of Physics, Kyushu University, Fukuoka 812-8581, Japan}

\date{\today}

\begin{abstract}
We analyze recently-measured total reaction cross sections
for $^{24\mbox{--}38}$Mg isotopes incident
on $^{12}$C targets at 240 MeV/nucleon
by using the folding model and antisymmetrized molecular dynamics (AMD).
The folding model well reproduces the measured reaction cross sections,
when the projectile densities are evaluated by
the deformed Woods-Saxon (def-WS) model with AMD deformation.
Matter radii of $^{24\mbox{--}38}$Mg are then deduced from
the measured reaction cross sections by fine-tuning the parameters of
the def-WS model. The deduced matter radii are largely enhanced
by nuclear deformation.
Fully-microscopic AMD calculations with no free parameter
well reproduce the deduced matter radii for $^{24\mbox{--}36}$Mg,
but still considerably underestimate them for $^{37,38}$Mg.
The large matter radii suggest that
$^{37,38}$Mg are candidates for deformed halo nucleus.
AMD also reproduces other existing measured ground-state properties
(spin-parity, total binding energy, and one-neutron separation energy) of
Mg isotopes.
Neutron-number ($N$) dependence of deformation parameter is predicted by AMD.
Large deformation is seen from $^{31}$Mg with $N=19$ to a drip-line nucleus
$^{40}$Mg with $N=28$,
indicating that both the $N=20$ and 28 magicities disappear.
$N$ dependence of neutron skin thickness is also predicted by AMD.
\end{abstract}


\maketitle
\section{Introduction}
Elucidation of properties of unstable nuclei is
an important subject in nuclear physics.
Some exotic properties were found so far in unstable nuclei particularly near
the drip line.
One is the change of well-known magicity.
The standard shell ordering can evolve as a function of neutron number $N$
or proton number $Z$ due to residual nucleon-nucleon interactions. This leads
to a quenching of some shell gap and consequently the change of magic numbers.
When the abrupt onset of the change of magic numbers appears in a region,
the region is called the ``Island of inversion".
In the region, nuclei have larger binding energies than expected
due to collectivity such as rotation.

As a pioneering work,
Klapisch and Thibault first revealed anomalies in the binding energies
of neutron-rich Na isotopes~\cite{Klapisch-1969,Thibault-1975},
and Warburton {\it et al.} predicted that
$(sd)^{-2}(fp)^2$ intruder configurations form the ground states in
the $N=20 \sim 22$ region of Ne, Na and Mg isotopes~\cite{Warburton}.
This prediction was supported by mass measurements~\cite{Orr-1991}.
In the region, strong deformation of nuclei was suggested by measured
low excitation energies and large $B(E2)$ values
of the first excited states~\cite{Mot95,Caurier,Utsuno,Iwas01,Yana03}.
As a mechanism behind the shell-gap quenching, more recently,
the importance of the nucleon-nucleon tensor interaction was pointed
out by Otsuka {\it et al.}~\cite{Otsuka-2005,Otsuka-2010}.
The $N=20$ magicity is thus considered to disappear in Ne, Na and Mg isotopes.

Another important progress in physics of unstable nuclei
is the discovery of halo nuclei
by measurements of interaction cross sections $\sigma_{\rm I}$
or one- and two-neutron removal cross sections~\cite{Tanihata,Jensen,Jonson}.
Here $\sigma_{\rm I}$ is used as a substitute of reaction cross sections
$\sigma_{\rm R}$, since the two cross sections are nearly identical
for the scattering of
unstable nuclei at intermediate and high incident energies~\cite{Sumi:2012}.
Now, $^{6}$He, $^{11}$Li, $^{11}$Be and others are considered
to be halo nuclei.
Recently, Nakamura {\it et al.}~\cite{Nakamura} suggested
through measurements of the one-neutron removal cross section of $^{31}$Ne
on C and Pb targets at 240~MeV/nucleon
that $^{31}$Ne is a halo nucleus which resides in the island of inversion.
Takechi {\it et al.}~\cite{Takechi} measured $\sigma_{\rm I}$
for Ne isotopes incident on $^{12}$C targets at 240~MeV/nucleon and
came to the same conclusion as Nakamura {\it et al}.
Very recently, Takechi {\it et al.} measured $\sigma_{\rm R}$
for $^{24\mbox{--}38}$Mg isotopes on C targets
at 240~MeV/nucleon~\cite{Takechi-Mg} and
suggested that $^{37}$Mg is a halo nucleus.

A powerful tool of analyzing measured $\sigma_{\rm R}$ or $\sigma_{\rm I}$
microscopically is
the folding model with
the $g$-matrix effective nucleon-nucleon
interaction~\cite{M3Y,JLM,Brieva-Rook,Satchler-1979,Satchler,CEG,
Rikus-von Geramb,Amos,CEG07,rainbow,Toyokawa:2013uua}.
For nucleon scattering from stable target nuclei,
the folding potential with the Melbourne $g$-matrix interaction
reproduces measured elastic and reaction cross sections
systematically with no adjustable parameter~\cite{Amos,Toyokawa:2013uua}.
The folding model is reliable also
for the scattering of unstable nuclei from stable target nuclei
at intermediate incident energies, say 200~MeV/nucleon, since
the projectile breakup is small there.
In fact, for $^{31}$Ne scattering from $^{12}$C targets
at 240~MeV/nucleon, the breakup cross section is about 1\% of
$\sigma_{\rm R}$~\cite{ERT}. This indicates
that the folding model is reliable also for other nucleus-nucleus scattering
at intermediate energies, since $^{31}$Ne is one of the most weakly bound
systems.

The optical potential of nucleus-nucleus scattering
is obtained by folding the $g$-matrix with the projectile and target densities.
When the projectile is deformed, the density profile is elongated by the
deformation.
The elongation enlarges the root-mean-square (rms) radius of the projectile
and eventually $\sigma_{\rm R}$. Recently,
antisymmetrized molecular dynamics (AMD) with
the Gogny-D1S interaction~\cite{GognyD1S} was applied to nuclei
in the island of inversion~\cite{Kimura,Kimura1}.
The calculations with the angular momentum projection (AMP) show
that the nuclei are largely deformed. The result is consistent
with that of the Hartree-Fock-Bogoliubov calculations
with the AMP~\cite{RER02,RER03}.
Here $n$-particle $m$-hole excitations in the Nilsson orbitals are essential
to determine the deformed configurations.

Very recently, the folding model with AMD projectile density
succeeded in reproducing measured $\sigma_{\rm I}$ for $^{28\mbox{--}30,32}$Ne
in virtue of large deformation of the projectiles~\cite{Minomo:2011bb}.
For $^{31}$Ne, the theoretical calculation underestimated measured
$\sigma_{\rm I}$ by about 3\%.
The AMD density has an inaccurate asymptotic form, since
each nucleon is described by a Gaussian wave packet in AMD.
The error coming from the inaccuracy is not negligible when
the one-neutron separation energy $S_n$ is small, say \mbox{$S_n \la 1$~MeV},
and thereby the nucleus has halo structure~\cite{Sumi:2012}.
The tail correction to the AMD density was then made by
the $^{30}$Ne + $n$ resonating group method (RGM) with core excitations in which
the deformed ground and excited states of $^{30}$Ne are calculated
with AMD~\cite{Minomo:2011bb}.
The folding model with the tail-corrected AMD-RGM density reproduces measured
$\sigma_{\rm I}$ for $^{31}$Ne. The fact leads to the conclusion that
$^{31}$Ne is the deformed halo nucleus which resides
in the island of inversion.

The deformed Woods-Saxon (def-WS) model with AMD deformation
provides the matter density with the proper asymptotic form~ \cite{Minomo-DWS}.
The def-WS model well reproduces measured $\sigma_{\rm R}$
for $^{20\mbox{--}32}$Ne, and the results of the def-WS density with AMD deformation
are consistent with those of the AMD density for $^{20\mbox{--}30,32}$Ne and with
that of the AMD-RGM density for $^{31}$Ne.
The def-WS model with AMD deformation is thus a handy way
of simulating AMD or AMD-RGM densities.
The def-WS model with AMD deformation also reproduces
measured $\sigma_{\rm R}$ for $^{24\mbox{--}36}$Mg~\cite{Takechi-Mg}.
As another advantage of the def-WS model,
one can fine-tune the theoretical result to the experimental data precisely
by changing the potential parameters or the deformation parameter slightly.
Making this analysis for $^{37}$Mg,
we suggested in our previous work~\cite{Takechi-Mg}
that $^{37}$Mg is a deformed halo nucleus.

In this paper, we first determine matter radii of $^{24\mbox{--}38}$Mg
systematically from measured $\sigma_{\rm R}$
by fine-tuning the parameters of the def-WS model.
This flexibility is an advantage of the def-WS model.
We next confirm that matter radii are largely enhanced
by nuclear deformation for Mg isotopes.

Fully microscopic AMD calculations, meanwhile, have no adjustable parameter
and hence high predictability, if the results of AMD calculations are
consistent with existing experimental data.
We then compare the deduced matter radii with the results of
AMD calculations.
The calculations are successful in reproducing the deduced
matter radii particularly for $^{24\mbox{--}36}$Mg.
For $^{37,38}$Mg, meanwhile,
AMD calculations considerably underestimate the deduced matter radii.
This suggests that $^{37,38}$Mg are candidates for deformed halo nucleus.
AMD calculations are also successful in reproducing existing
experimental data on other ground-state properties
(spin-parity, total binding energy, and one-neutron separation energy)
of Mg isotopes. $N$ dependence of deformation parameter and
neutron-skin thickness is therefore predicted with AMD.

The theoretical framework is presented in Sec.~\ref{Theoretical framework}.
Numerical results are shown in Sec.~\ref{Results}.
Comparison of the def-WS results with the AMD results is made and
structure of $^{37}$Mg is discussed in Sec.~\ref{Discussions}.
Section~\ref{Summary} is devoted to a summary.

\section{Theoretical framework}
\label{Theoretical framework}

In this section, we recapitulate the folding model, AMD and the def-WS model.
Further explanation is presented in Ref.~\cite{Sumi:2012}
for the folding and def-WS models and Refs.~\cite{Kimura1,Sumi:2012} for AMD.

\subsection{Folding model}

The nucleus-nucleus scattering is governed by
the many-body Schr\"odinger equation,
\bea
(T_R +h_{\rm P}+h_{\rm T}+ \sum_{i \in {\rm P}, j
\in {\rm T}} v_{ij}-E){\Psi}^{(+)}=0  \;,
\label{schrodinger-bare}
\eea
with the realistic nucleon-nucleon interaction $v_{ij}$,
where
$E$ is the total energy, $T_R$ is the kinetic energy between
the projectile (P) and the target (T), and
$h_{\rm P}$ ($h_{\rm T}$) is the internal Hamiltonian of P (T).
Equation~\eqref{schrodinger-bare} is reduced to
\bea
(T_R +h_{\rm P}+h_{\rm T}+ \sum_{i \in {\rm P}, j \in {\rm T}}
\tau_{ij}-E){\hat \Psi}^{(+)}=0
\label{schrodinger-effective}
\eea
by using the multiple-scattering theory~\cite{Watson, KMT} for
nucleus-nucleus scattering~\cite{Yahiro-Glauber}.
Here $\tau_{ij}$ is the effective nucleon-nucleon interaction
in the nuclear medium, and the $g$-matrix is often used
as such $\tau_{ij}$~\cite{M3Y,JLM,Brieva-Rook,Satchler-1979,Satchler,CEG,
Rikus-von Geramb,Amos,CEG07,rainbow}.
At intermediate and high incident energies of our interest,
breakup and collective excitations of P and T are small~\cite{Sumi:2012}, and
Eq.~\eqref{schrodinger-effective} is further reduced to
the single-channel equation
\bea
[T_R+U-E_{\rm in}]\psi=0,
\eea
with the folding potential
\bea
U=\langle \Phi_0 | \sum_{i \in {\rm P}, j \in {\rm T}}
\tau_{ij} | \Phi_0 \rangle \;,
\eea
where $E_{\rm in}$ is the incident energy,
$\Phi_0$ is the product of the ground states of P and T, and
$\psi$ is the relative wave function between P and T.
This is nothing but the folding model based on the $g$-matrix interaction.
In particular, the folding model with the Melbourne $g$-matrix~\cite{Amos}
well reproduces $\sigma_{\rm R}$ for
Ne isotopes incident on a $^{12}$C target
at 240 MeV/A~\cite{Minomo-DWS,Minomo:2011bb,Sumi:2012}.

The potential is composed of the direct and exchange parts,
$U^{\rm DR}$ and $U^{\rm EX}$~\cite{DFM-standard-form,DFM-standard-form-2}:
\bea
\label{eq:UD}
U^{\rm DR}(\vR) \hspace*{-0.15cm} &=& \hspace*{-0.15cm}
\sum_{\mu,\nu}\int \rho^{\mu}_{\rm P}(\vrr_{\rm P})
            \rho^{\nu}_{\rm T}(\vrr_{\rm T})
            g^{\rm DR}_{\mu\nu}(s;\rho_{\mu\nu}) d \vrr_{\rm P} d \vrr_{\rm T}, \\
\label{eq:UEX}
U^{\rm EX}(\vR) \hspace*{-0.15cm} &=& \hspace*{-0.15cm}\sum_{\mu,\nu}
\int \rho^{\mu}_{\rm P}(\vrr_{\rm P},\vrr_{\rm P}-\vs)
\rho^{\nu}_{\rm T}(\vrr_{\rm T},\vrr_{\rm T}+\vs) \nonumber \\
            &&~~\hspace*{-0.5cm}\times g^{\rm EX}_{\mu\nu}(s;\rho_{\mu\nu}) \exp{[-i\vK(\vR) \cdot \vs/M]}
            d \vrr_{\rm P} d \vrr_{\rm T},~~~~
            \label{U-EX}
\eea
where $\vR$ is the relative coordinate between P and T,
$\vs=\vrr_{\rm P}-\vrr_{\rm T}+\vR$,
and $\vrr_{\rm P}$ ($\vrr_{\rm T}$) is
the coordinate of the interacting nucleon from P (T).
Each of $\mu$ and $\nu$ denotes the $z$-component of isospin; 1/2 means neutron and $-1/2$ does proton.
The nonlocal $U^{\rm EX}$ has been localized in Eq.~\eqref{U-EX}
with the local semi-classical approximation~\cite{Brieva-Rook},
where \vK(\vR) is the local momentum between P and T, and
$M=A A_{\rm T}/(A +A_{\rm T})$
for the mass number $A$ ($A_{\rm T}$) of P (T);
see Refs.~\cite{Hag06,Minomo:2009ds} for the validity of the localization.
The direct and exchange parts, $g^{\rm DR}_{\mu\nu}$ and
$g^{\rm EX}_{\mu\nu}$, of the $g$-matrix depend on the local density
\bea
 \rho_{\mu\nu}=\rho^{\mu}_{\rm P}(\vrr_{\rm P}-\vs/2)
 +\rho^{\nu}_{\rm T}(\vrr_{\rm T}+\vs/2)
\label{local-density approximation}
\eea
at the midpoint of the interacting nucleon pair; see
Ref.~\cite{Sumi:2012} for the explicit forms of $g^{\rm DR}_{\mu\nu}$ and
$g^{\rm EX}_{\mu\nu}$.

The relative wave function $\psi$ is decomposed into partial waves $\chi_L$,
each with different orbital angular momentum $L$.
The elastic $S$-matrix elements $S_L$ are obtained from the asymptotic form of
the $\chi_L$.
The total reaction cross section $\sigma_{\rm R}$ is calculable from
the $S_L$ as
\bea
\sigma_{\rm R}=\frac{\pi}{K^2}\sum_L (2L+1)(1-|S_L|^2).
\eea
The potential $U$ has the nonspherical part when the projectile spin is
nonzero.  In the present calculation, we neglect the nonspherical part
since the approximation is confirmed to be quite good
for the reaction cross section~\cite{Sumi:2012,Toyokawa:2013uua}.

\subsection{AMD}

AMD starts with the many-body Schr\"odinger equation
\begin{align}
 H&= T + \sum_{i<j} {\bar v}_{ij} - T_{\rm cm}
\end{align}
with the effective nucleon-nucleon interaction ${\bar v}_{ij}$,
where the center-of-mass kinetic energy $T_{\rm cm}$ is subtracted from
the kinetic energy $T$ of nucleons.
In this paper, we use the Gogny-D1S effective nucleon-nucleon interaction~\cite{GognyD1S}
plus Coulomb interaction as ${\bar v}_{ij}$.

The variational wave function $\Phi^\pi$ is
parity projected from
the Slater determinant of nucleon wave packets:
\begin{align}
 \Phi^\pi&=P^\pi{\cal A}\left\{\varphi_1,\varphi_2,...,\varphi_A \right\} \;,
\label{eq:amdint}
\end{align}
where $P^\pi$ (${\cal A}$) is the parity-projection (antisymmetrization)
operator. The $i$th single-particle wave packet $\varphi_i$ is defined
by $\varphi_i=\phi_i(\bm r)\chi_i\xi_i$ with
\begin{align}
 \phi_i(\bm r) &= \prod_{\sigma=x,y,z}\left(\frac{2\nu_\sigma}{\pi}\right)^{1/4}
 \exp\left\{-\nu_\sigma\left(r_{\sigma} - \frac{Z_{i\sigma}}{\sqrt{\nu_\sigma}}\right)^2\right\}, \\
 \chi_i &= \alpha_{i,\uparrow} \chi_{\uparrow} + \alpha_{i,\downarrow}
 \chi_{\downarrow}, \hspace{5mm}|\alpha_{i,\uparrow}|^2+|\alpha_{i,\downarrow}|^2=1,\\
 \quad \xi_i &= \text{$p$ or $n$}.
\end{align}
The centroids $\bm Z_i$, the width $\nu_\sigma$
and the spin directions $\alpha_{i,\uparrow}$ and $\alpha_{i,\downarrow}$ of
Gaussian wave packets are variational parameters.
The center-of-mass wave function
can be analytically separated
from the variational wave function.
Hence all quantities calculated with AMD are free
from the spurious center-of-mass motion.

The parameters in Eq.~\eqref{eq:amdint} are determined
with the frictional cooling method by
minimizing the total energy under the constraint
on the matter quadrupole deformation parameter $\bar{\beta}$ defined by
\begin{align}
 \frac{\left< x^2 \right>^{1/2}}
 {\left[ \left< x^2 \right> \left< y^2 \right> \left< z^2 \right>
 \right]^{1/6}} &= \exp \left[ {\sqrt {\frac{5}{4\pi}}} \bar{\beta} \cos
 \left( \bar{\gamma} + \frac{2\pi}{3} \right) \right],
 \label{eq:AMDbgx2}\\
 \frac{\left< y^2 \right>^{1/2}}
 {\left[ \left< x^2 \right> \left< y^2 \right> \left< z^2 \right>
 \right]^{1/6}} &= \exp \left[ {\sqrt {\frac{5}{4\pi}}} \bar{\beta} \cos
 \left( \bar{\gamma} - \frac{2\pi}{3} \right) \right],
 \label{eq:AMDbgy2}\\
 \frac{\left< z^2 \right>^{1/2}}
 {\left[ \left< x^2 \right> \left< y^2 \right> \left< z^2 \right>
 \right]^{1/6}} &= \exp \left[ {\sqrt {\frac{5}{4\pi}}} \bar{\beta} \cos
 \bar{\gamma}  \right].
 \label{eq:AMDbgz2}
\end{align}
Here, $\left< x^2 \right>$, $\left< y^2 \right>$ and
$\left< z^2 \right>$ defined in the  intrinsic frame are so chosen
to satisfy the ordering
$\left< x^2 \right> \le \left< y^2 \right> \le\left< z^2 \right>$.
Since no constraint is imposed on $\bar{\gamma}$, it has an
optimal value for each value of $\bar{\beta}$.

After the variation, we perform the AMP for each value of $\bar{\beta}$,
\begin{align}
 \Phi^{I\pi}_{mK}(\bar{\beta})&=P^{I}_{mK}\Phi_{\rm int}^\pi(\bar{\beta}),\label{eq:amdproj}\\
 P^{I}_{mK} &= \frac{2I+1}{8\pi^2}\int d\Omega D^{I*}_{mK}(\Omega)R(\Omega),\label{eq:jproj}
\end{align}
where $D^I_{mK}(\Omega)$ and $R(\Omega)$ are the Wigner's $D$ function and
the rotation operator, respectively.
The wave functions that have the same
parity and angular momentum $(I, m)$ are superposed as
\begin{align}
 \Phi_n^{Im\pi}&= \sum_{K=-I}^{I} \sum_{\bar{\beta}} c_{nK} (\bar{\beta})
 \Phi_{mK}^{I\pi}(\bar{\beta}) ,
\label{eq:AMD-wf}
\end{align}
where $\bar{\beta}$ is varied from 0 to 1 with an interval of 0.025
in actual calculations.
The coefficients $c_{nK}(\bar{\beta})$ are determined by solving
the Hill-Wheeler equation.

The ground state wave function $\Phi^{Im\pi}_{\rm g.s.}$ thus obtained is
transformed into the nucleon density as
\begin{align}
 \rho_{ImIm'}(\bm r)&=\langle \Phi^{Im\pi}_{\rm g.s.}|\sum_{i}
 \delta(\bm r_i - \bm X - \bm r)|\Phi^{Im'\pi}_{\rm g.s.}\rangle,
 \label{eq:amddens1}\\
 &=\sum_{\lambda=0}^{2I} \rho_{II}^{(\lambda)}(r)
 (Im' \lambda \mu |Im)
 Y^*_{\lambda \mu}(\hat {\bm r}),
 \label{eq:amddens}
\end{align}
where ${\bm X}$ denotes the center-of-mass coordinate.
When $I > 0$, the multipolarity $\lambda$ can take nonzero values.
The nonzero components
make the folding potential $U$ nonspherical.
But the effects are small on $\sigma_{\rm R}$~\cite{Sumi:2012,Toyokawa:2013uua}.
We then take only the spherical ($\lambda=0$) component
in this paper.

\subsection{def-WS model}

The def-WS potential consists of the central and spin-orbit parts:
\begin{align}
\label{DWS-0}
   V_{\rm c}( {\bfi r})
 = \frac{V_0}{1+\exp \left[ {\rm dist}_{\Sigma}
   ({\bfi r})/a \right]},
\end{align}
\begin{align}
\label{DWS-1}
   V_{\rm so}( {\bfi r})
 &= \lambda_{\rm so}\left(\frac{\hbar}{2m_{\rm red}c}\right)^2
  \bm{\nabla}V_{\rm c}(\bm{r})\cdot
  \left(\bm{\sigma}\times \frac{1}{i}\bm{\nabla}\right),
\end{align}
where
$m_{\rm red}=m(A-1)/{A}$ for nucleon mass $m$
and the function ${\rm dist}_{\Sigma} ({\bfi r})$
represents the distance of a point ${\bfi r}$ from
the deformed surface $\Sigma$ that is specified by the radius
\begin{align}
 R(\theta,\phi;\bm{\alpha}) = R_{0}c_{v}(\bm{\alpha})\Bigr[
 1+\sum_{\lambda\mu}\alpha^*_{\lambda\mu}Y_{\lambda\mu}(\theta,\phi)\Bigl],
\label{eq:surf}
\end{align}
with the deformation parameters
$\bm{\alpha} \equiv \left\{ \alpha_{\lambda\mu}\right\} $.
The constant $c_v (\bm{\alpha})$ is introduced to guarantee
the volume conservation of nucleus.
Since the effect of the hexadecapole deformation on the reaction cross section
is rather small~\cite{Sumi:2012}, we only include the quadrupole deformation
in this study.
The parameter set $(\alpha_{2\mu})$ is related to
the standard set $(\beta_2,\gamma)$ as
\begin{equation}
\left\{ \begin{array}{l}
 \alpha_{20}=\beta_2\cos\gamma,\\
 \alpha_{22}=\alpha_{2-2}=-\frac{1}{\sqrt{2}}\,\beta_2\sin\gamma,\\
\end{array} \right.
\label{eq:WSbeta}
\end{equation}
As the parameter set $(V_0,R_0,a,\lambda_{\rm so})$
of the Woods-Saxon (WS) potential,
we take a recent parametrization given by R.~Wyss~\cite{WyssPriv};
see Appendix~\ref{Parameter set of Woods-Saxon potential} for actual values
of the parameter set.

The deformation parameters ($\beta_2$, $\gamma$) in the def-WS model are  determined from the corresponding deformation parameters 
($\bar{\beta}$, $\bar{\gamma}$) in AMD.
 Here, the parameters ($\bar{\beta}$, $\bar{\gamma}$) of 
 AMD wave function $\Phi_n^{Im\pi}$ (Eq. (\ref{eq:AMD-wf})) are defined as 
those of the basis wave function $\Phi^{I\pi}_{mK}(\bar{\beta})$ 
(Eq. (\ref{eq:amdproj})) that has the maximum overlap with $\Phi_n^{Im\pi}$.
The relation between $(\bar{\beta},\bar{\gamma})$
in AMD and $(\beta_2,\gamma)$ in the def-WS model is obtained
so that both the models can yield the same ratio
$\langle x^2 \rangle:\langle y^2 \rangle:\langle z^2 \rangle$;
see Appendix~\ref{Relation of defromation parameters} for further explanation
and actual values of $\bar{\beta}$ and $\beta_2$ for Mg isotopes.

The nucleon density calculated by the def-WS model is the intrinsic density
in the body-fixed frame and unisotropic,
while the density used in the folding model is that in the laboratory frame.
In order to obtain the latter from the former one has to perform the AMP.
Instead, we use the angle-average of the deformed intrinsic density,
which has been confirmed to be a good approximation
of the projected density; see Ref.~\cite{Sumi:2012} for details.

\subsection{Spherical HF and HFB}
As a reference, the spherical Hartree-Fock (HF) and
spherical Hartree-Fock-Bogoliubov (HFB) methods are employed
to calculate the nucleon density for the spherical systems.
In the HF and HFB calculations it is important
to properly choose the effective interaction in order to obtain
accurate description of the ground state.
Ee use consistently the same Gogny-D1S interaction~\cite{GognyD1S}
as in the AMD calculation for this purpose.
We refer to the spherical Gogny HF and HFB methods
as sph-GHF and sph-GHFB in this paper.
The spherical shape is imposed with
the filling approximation as a standard manner,
whenever sph-GHF and sph-GHFB calculations are done.
In actual sph-GHF and sph-GHFB calculations, we adopt the
Gaussian expansion method~\cite{Nakada:2008} that reduces numerical tasks.

\section{Results}
\label{Results}

\subsection{Reaction cross sections for stable nuclei}

First, the accuracy of the Melbourne $g$-matrix folding model
(i.e., the folding model with the Melbourne $g$-matrix interaction)
is tested for the scattering of $^{12}$C from
several stable nuclei at intermediate energies around 250 MeV/nucleon.
The $g$-matrix folding model was successful in reproducing
measured reaction cross sections for $^{12}$C, $^{20}$Ne, $^{23}$Na
and $^{27}$Al~\cite{Minomo-DWS,Minomo:2011bb,Sumi:2012}.
In this paper, we newly consider $^{24}$Mg and $^{40}$Ca targets
in addition to the stable targets mentioned above.
We employ the phenomenological densities~\cite{C12-density}
as the projectile and target densities,
where the proton density is deduced from the electron scattering and
the neutron distribution is assumed to have the same geometry
as the proton one. This assumption is good for the stable nuclei,
since the neutron rms radii are almost the same as the proton ones
in sph-GHF calculations.

\begin{figure}[htbp]
\begin{center}
 \includegraphics[width=0.4\textwidth,clip]{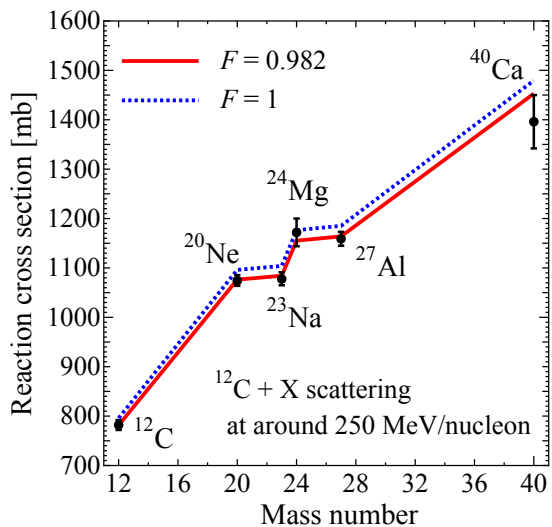}
 \caption{(Color online) Reaction cross sections for the $^{12}$C scattering
 on stable nuclei from $A=12$ to 40. The data for $^{12}$C and $^{27}$Al
 at $250.8$~MeV/nucleon are taken from Ref.~\cite{expC12C12}.
 The data for $^{20}$Ne and $^{23}$Na at $250$~MeV/nucleon are
 deduced from the measured $\sigma_{\rm I}$
 at around 1~GeV/nucleon~\cite{Ne20-sigmaI,Na23-sigmaI}
 with the Glauber model~\cite{Takechi}.
 The data for $^{40}$Ca at $240$~MeV/nucleon is obtained
 from the measured $\sigma_{\rm R}$
 at 83~MeV/nucleon~\cite{Kox-Ca-1984}
 with the Glauber model~\cite{Takechi-Mg}.
 The solid (dotted) line stands for the results of the folding-model 
 calculations after (before) the normalization with a factor $F=0.982$.
}
 \label{Fig-reaction-Xsec-stable}
\end{center}
\end{figure}

Figure~\ref{Fig-reaction-Xsec-stable} shows reaction cross sections
for the $^{12}$C scattering from $^{12}$C, $^{20}$Ne, $^{23}$Na, $^{24}$Mg, $^{27}$Al and $^{40}$Ca targets at around 250 MeV/nucleon.
The experimental data are taken
from Refs.~\cite{expC12C12,Ne20-sigmaI,Na23-sigmaI,Kox-Ca-1984,Takechi-Mg}.
The results of the folding-model (dotted line)
well reproduce the experimental data.
More precisely, they slightly overestimate the data.
We then introduce the normalization factor of
$F=0.982$ to reproduce the mean value of the experimental
data for $^{12}$C. The result (solid line) is consistent with
experimental data for other targets.
This fine tuning is taken also for the scattering of Mg isotopes
from $^{12}$C targets around 240~MeV/nucleon.

\subsection{Matter radii of Mg isotopes}
\label{Matter radii of Mg isotopes}

\begin{figure}[htbp]
\begin{center}
\hspace*{-0.5cm}
 \includegraphics[width=0.4\textwidth,clip]{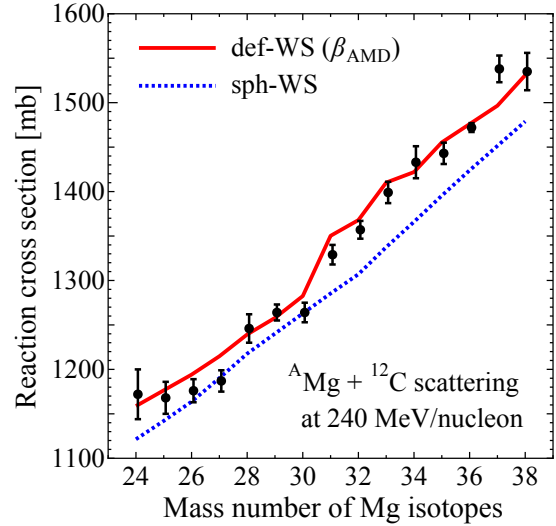}
 \caption{(Color online)
Reaction cross sections for the scattering of Mg isotopes at 240 MeV/nucleon.
The experimental data are taken from Ref.~\cite{Takechi-Mg}.
The solid line stands for the results of the def-WS model with AMD deformation,
whereas the dotted line corresponds to the results of spherical Woods-Saxon (sph-WS) calculations.
 }
 \label{RCS_Mg_DWS-SWS}
\end{center}
\end{figure}

Figure~\ref{RCS_Mg_DWS-SWS} shows
a comparison of calculated and measured $\sigma_{\rm R}$
for the scattering of $^{24\mbox{--}38}$Mg on a $^{12}$C target at 240~MeV/nucleon.
The $\sigma_{\rm R}$ are evaluated by the folding model with
different types of projectile densities; one is the densities
calculated by the def-WS model with AMD deformation and
the other is the densities of spherical Woods-Saxon (sph-WS) calculations.
The solid line denotes the results of the def-WS model,
whereas the dotted line corresponds to the results of sph-WS calculations.
The large difference between the two lines shows that
nuclear deformation effects are important in $\sigma_{\rm R}$.
The def-WS model yields good agreement with the experimental
data~\cite{Takechi-Mg}.
For $^{37}$Mg, however, the def-WS model slightly underestimates
the measured $\sigma_{\rm R}$, indicating that
$^{37}$Mg is a deformed halo nucleus; see Ref.~\cite{Takechi-Mg}
and Sec.~\ref{Sec:37Mg} for the detail.

Since the def-WS results are consistent with the measured $\sigma_\mathrm{R}$,
we deduce the rms radii of Mg isotopes from
the measured $\sigma_\mathrm{R}$ by fitting it with
the calculation, where either the depth parameter
$V_0$ or the deformation parameter $\beta_2$ are adjusted
slightly in the def-WS potential.
This flexibility is a merit of the def-WS model.

\begin{figure}[htbp]
\begin{center}
 \includegraphics[width=0.4\textwidth,clip]{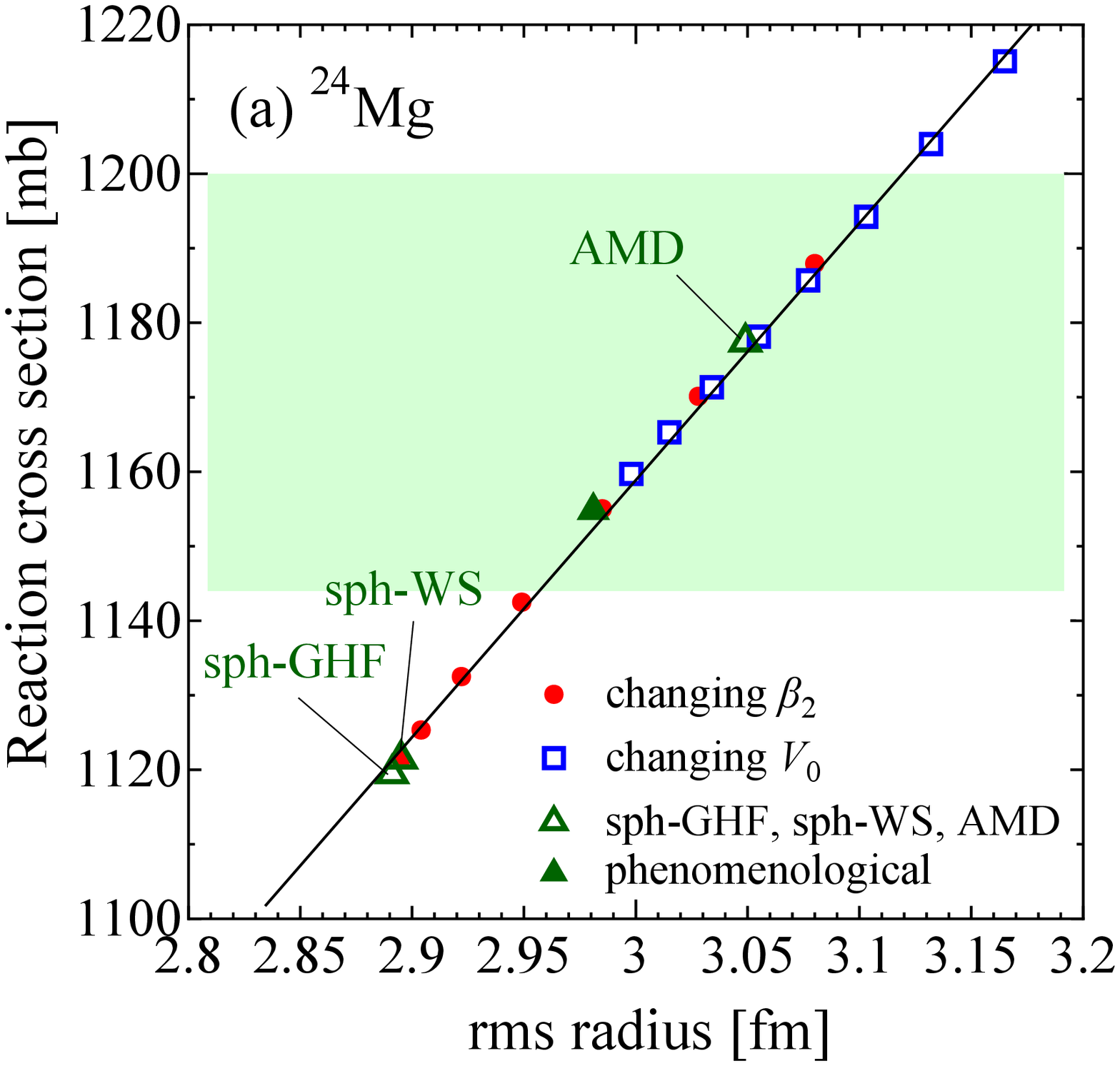}
 \includegraphics[width=0.4\textwidth,clip]{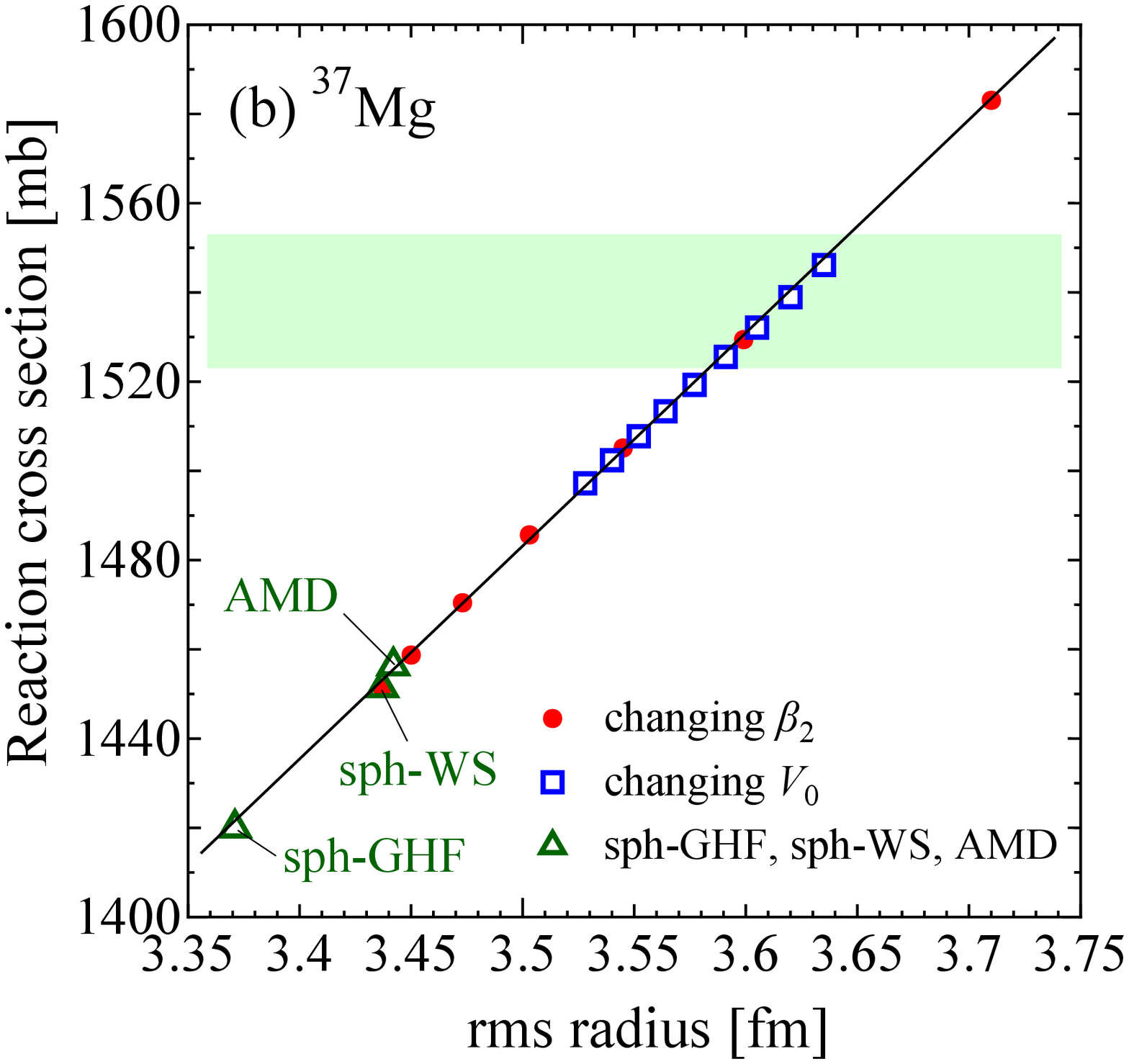}
 \caption{(Color online)
Relation between the rms matter radius and the reaction cross section
for (a) $^{24}$Mg and (b) $^{37}$Mg.
The closed circles stands for the results of the def-WS model
in which $\beta_2$ is varied from 0 to 0.6 with interval 0.1, and
the open squares correspond to the results of the def-WS model in which $|V_0|$
is reduced by a factor $1 \sim 0.72$ with interval 0.04
for $^{24}$Mg in panel (a) and by a factor $1 \sim 0.92$ 
with interval 0.01 for $^{37}$Mg in panel (b). 
Smaller $\beta_2$ corresponds to smaller rms matter radius and hence smaller 
reaction cross section, while smaller $|V_0|$ does to larger rms matter radius 
and hence larger reaction cross section. 
The open triangles denote the results of sph-GHF, sph-WS,
AMD densities from bottom. For $^{24}$Mg,
the result of the phenomenological density~\cite{C12-density}
is also presented by a closed triangle.
The hatching region shows the lower and upper bounds
of measured $\sigma_{\rm R}$.
}
 \label{RCS-r}
\end{center}
\end{figure}

The relation between $\sigma_\mathrm{R}$ and the corresponding rms matter
radius is plotted in Fig.~\ref{RCS-r}
for two  nuclei; (a) $^{24}$Mg and (b) $^{37}$Mg.
The closed circles denote the results of the def-WS model in which $\beta_2$ is
varied from 0 to 0.6 with the interval of 0.1
with keeping all other parameters.
Note that the AMD deformation used in Fig.~\ref{RCS_Mg_DWS-SWS}
is $\beta_2=0.433$ for $^{24}$Mg and is $\beta_2=0.362$ for $^{37}$Mg,
respectively (see Table~\ref{tab:DWS-parameter}).
Larger $\beta_2$ yields
larger rms radius and hence larger $\sigma_\mathrm{R}$.
The open squares correspond to the results of the def-WS model in which $|V_0|$
is reduced by a factor $1 \sim 0.72$ with interval 0.04
for $^{24}$Mg in panel (a) and by a factor $1 \sim 0.92$ 
with interval 0.01 for $^{37}$Mg in panel (b).
The results of sph-GHF, spherical Woods-Saxon (sph-WS),
AMD densities are also shown
by open triangles from bottom (i.e., they are in increasing order).
For $^{24}$Mg,
the result of the phenomenological density deduced from
 the electron scattering~\cite{C12-density} is also presented by a closed triangle.
As an important result,
all the results are on a straight line
for each case of (a) $^{24}$Mg and of (b) $^{37}$Mg.
We can then precisely determine matter radii of Mg isotopes
corresponding to the measured $\sigma_\mathrm{R}$
by using the straight line.

\begin{figure}[htbp]
\begin{center}
 \includegraphics[width=0.40\textwidth,clip]{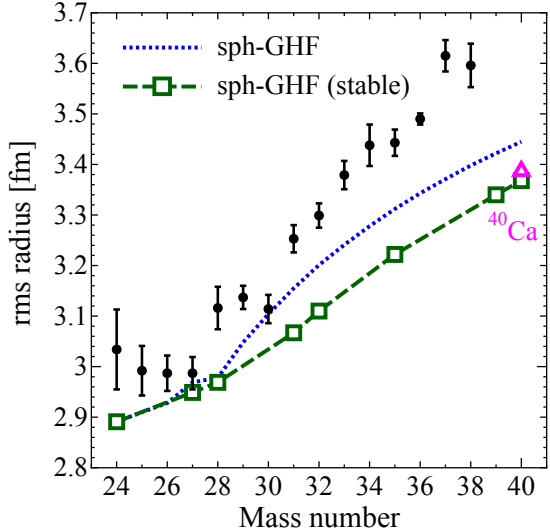}
 \caption{(Color online)
Matter radii of Mg-isotopes deduced from
measured $\sigma_\mathrm{R}$.
The dotted line denotes the results of sph-GHF calculations.
The calculated radii by the sph-GHF model for stable $A=24\mbox{--}40$ nuclei
are also included as the dashed line, where the nuclei included are
$^{24}$Mg, $^{27}$Al, $^{28}$Si,
$^{31}$P, $^{32}$S, $^{35}$Cl, $^{39}$K, and $^{40}$Ca.
For spherical nucleus $^{40}$Ca,
matter radius determined from the electron scattering~\cite{C12-density}
is shown by an open triangle.
}
 \label{rms_exp}
\end{center}
\end{figure}

The resultant matter radii of Mg isotopes are tabulated
in Table~\ref{tab:RMS_exp} and are plotted as a function of mass number $A$
in Fig.~\ref{rms_exp}.
The dotted and dashed lines represent the results of sph-GHF calculations
for Mg isotopes and stable $A=24\mbox{--}40$ nuclei, respectively.
The two lines correspond to matter radii of the nuclei in the spherical limit.
For the spherical nucleus $^{40}$Ca, the matter radius determined from
the electron scattering~\cite{C12-density} are plotted by an open triangle,
which lies on the dashed line,
indicating the reliability of sph-GHF calculations.
The difference between the dotted and dashed lines shows
the neutron-skin effects; the increase of neutron excess
makes the Fermi energy of neutrons much larger than that of protons,
and consequently the neutron radius
bulges out compared to those of stable $N\approx Z$ nuclei.
Large enhancement of the deduced radii from the dotted line mainly comes
from nuclear deformation. Further discussion is made in
the next Sec.~\ref{AMD analysis for Mg isotopes}.

For $^{37}$Mg, the matter radius should be carefully deduced from the measured 
$\sigma_{\rm R}$, since the def-WS model with AMD deformation considerably 
underestimates the measured $\sigma_{\rm R}$. 
In the previous analysis~\cite{Takechi-Mg}, we assumed that $^{37}$Mg is a 
deformed halo nucleus and reduced the potential depth only 
for the last neutron without changing the potential for the core nucleons. 
At $\beta_2=0.362$ calculated with AMD, 
the [312 5/2] orbital coming from the spherical 0$f_{7/2}$ orbital
has slightly lower energy
than the [321 1/2] orbital coming from the spherical $1p_{3/2}$ orbital. 
In the case that the last neutron is in the [312 5/2] orbital, 
the calculated $\sigma_{\rm R}$ cannot reproduce the measured $\sigma_{\rm R}$ 
even if the potential depth is reduced. 
This problem can be solved if the last neutron is in the [321 1/2] orbital. 
In the previous analysis, therefore, the last neutron is assumed to be 
in the [321 1/2] orbital and the potential depth is reduced only 
for the last neutron. 
In the present analysis, meanwhile, we simply consider that 
the last neutron is in the [312 5/2] orbital and reduce the potential depth 
uniformly for all the nucleons, i.e., 
the core ($^{36}$Mg) is also slightly expanded by the reduction. 
The deduced matter radius is $3.62\pm 0.03$~fm in the present
analysis and $3.65^{+0.09}_{-0.05}$~fm 
in the previous analysis, although in the previous analysis
the solid line shown in Fig.~\ref{RCS-r}(b) was slightly bended for 
Both are consistent with each other 
within the error bars. The matter radius is thus almost independent 
of the deduction procedure taken.

In Fig. 3(b), the matter radius of $^{37}$Mg calculated by the sph-WS model is considerably
larger than that by the sph-GHF model. This indicates that the depth of
the present parametrization of the WS potential is a bit too shallow 
for such an unstable nucleus with large neutron excess. This point will 
be discussed later in Sec.~\ref{Sec:DWS}.

\begin{table}
\caption{Matter radii of Mg isotopes deduced from measured $\sigma_\mathrm{R}$.
Mean values of deduced matter radii are evaluated from those of the corresponding $\sigma_\mathrm{R}$,
while errors of deduced matter radii are estimated from those of measured $\sigma_\mathrm{R}$
by the straight solid lines shown in Fig.~\ref{RCS-r}.
All the values are shown in units of fm.
}
\label{tab:RMS_exp}
\begin{center}
\begin{tabular}{cccc}\hline \hline
 nuclide   & rms radius & error \\ \hline
 $^{24}$Mg & 3.03 & 0.08 \\
 $^{25}$Mg & 2.99 & 0.05 \\
 $^{26}$Mg & 2.99 & 0.04 \\
 $^{27}$Mg & 2.99 & 0.03 \\
 $^{28}$Mg & 3.12 & 0.04 \\
 $^{29}$Mg & 3.14 & 0.02 \\
 $^{30}$Mg & 3.11 & 0.03 \\
 $^{31}$Mg & 3.25 & 0.03 \\
 $^{32}$Mg & 3.30 & 0.02 \\
 $^{33}$Mg & 3.38 & 0.03 \\
 $^{34}$Mg & 3.44 & 0.04 \\
 $^{35}$Mg & 3.44 & 0.03 \\
 $^{36}$Mg & 3.49 & 0.01 \\
 $^{37}$Mg & 3.62 & 0.03 \\
 $^{38}$Mg & 3.60 & 0.04 \\
 \hline
\end{tabular}
\end{center}
\end{table}

\subsection{AMD analyses for Mg isotopes}
\label{AMD analysis for Mg isotopes}

First, the ground-state properties, i.e.,
spin-parity $I^{\pi}$, one-neutron separation energy $S_{-1n}$
and deformation parameters $\bar{\beta}$ and $\bar{\gamma}$,
of Mg isotopes are predicted by AMD and tabulated in
Table~\ref{tab:AMD-result}.
AMD calculations yield the same $I^{\pi}$ as the data~\cite{Audi-2012}
for $^{24\mbox{--}34}$Mg.
For $^{35,37,39}$Mg, meanwhile, we cannot make definite discussion on
$I^{\pi}$, since it is not established experimentally.
For $^{37}$Mg, AMD calculations yield a small $S_{-1n}$ value consistent
with the empirical values $0.16\pm0.68$ MeV~\cite{Audi-2012},
though the error is large. For $^{39}$Mg,
 $S_{-1n}$ is negative in both the AMD calculation and
the empirical value~\cite{Audi-2012}.
In fact, $^{39}$Mg is experimentally shown to be unbound
in Ref.~\cite{Baumann-2007}.

\begin{table}
\caption{Ground-state properties of Mg isotopes predicted by AMD.
For $^{40}$Mg, the two-neutron separation energy $S_{-2n}$
is shown, since $^{39}$Mg is unbound in AMD calculations.
}
\label{tab:AMD-result}
\begin{center}
\begin{tabular}{cccccc}\hline \hline
 nuclide   & $I^{\pi}$(exp) & $I^{\pi}$(AMD) & $S_{-1n}$ [MeV]
 &  $\bar{\beta}$ & $\bar{\gamma}$  \\ \hline
 $^{24}$Mg & 0$^+$     & 0$^+$     &       & 0.42       &  0$^{^\circ}$  \\
 $^{25}$Mg & 5/2$^+$   & 5/2$^+$   & 7.125 & 0.40       &  0$^{^\circ}$  \\
 $^{26}$Mg & 0$^+$     & 0$^+$     &10.211 & 0.375      &  0$^{^\circ}$  \\
 $^{27}$Mg & 1/2$^+$   & 1/2$^+$   & 6.444 & 0.35       &  0$^{^\circ}$  \\
 $^{28}$Mg & 0$^+$     & 0$^+$     & 8.881 & 0.35       &  0$^{^\circ}$  \\
 $^{29}$Mg & 3/2$^+$   & 3/2$^+$   & 4.123 & 0.295      &  0$^{^\circ}$  \\
 $^{30}$Mg & 0$^+$     & 0$^+$     & 5.781 & 0.285      & 25$^{^\circ}$  \\
 $^{31}$Mg & 1/2$^+$   & 1/2$^+$   & 2.624 & 0.44       &  0$^{^\circ}$  \\
 $^{32}$Mg & 0$^+$     & 0$^+$     & 5.598 & 0.395      &  0$^{^\circ}$  \\
 $^{33}$Mg & 3/2$^-$   & 3/2$^-$   & 2.640 & 0.44       &  0$^{^\circ}$  \\
 $^{34}$Mg & 0$^+$     & 0$^+$     & 3.622 & 0.35       &  0$^{^\circ}$  \\
 $^{35}$Mg & (7/2$^-$) & 3/2$^+$   & 1.011 & 0.40       &  0$^{^\circ}$  \\
 $^{36}$Mg & 0$^+$     & 0$^+$     & 2.993 & 0.39       &  0$^{^\circ}$  \\
 $^{37}$Mg & (7/2$^-$) & 5/2$^-$   & 0.489 & 0.355      &  0$^{^\circ}$  \\
 $^{38}$Mg & 0$^+$     & 0$^+$     & 2.112 & 0.38       &  0$^{^\circ}$  \\
 $^{39}$Mg &           &           &unbound&            &                \\
 $^{40}$Mg & 0$^+$     & 0$^+$     & 1.119 ($S_{-2n}$) & 0.41       &  0$^{^\circ}$  \\
 \hline
\end{tabular}
\end{center}
\end{table}

\begin{figure}[htbp]
\begin{center}
 \includegraphics[width=0.40\textwidth,clip]{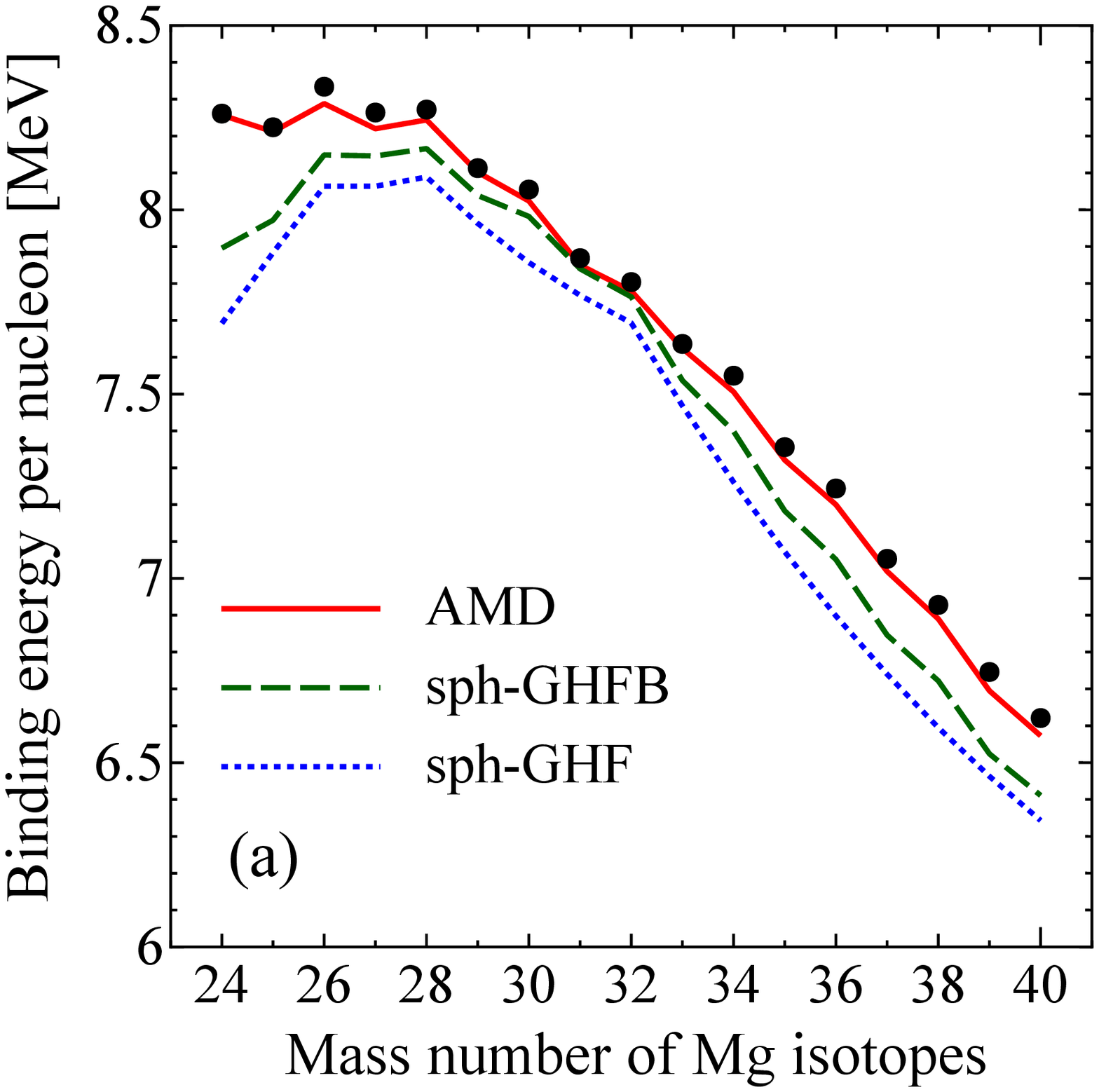}
 \includegraphics[width=0.40\textwidth,clip]{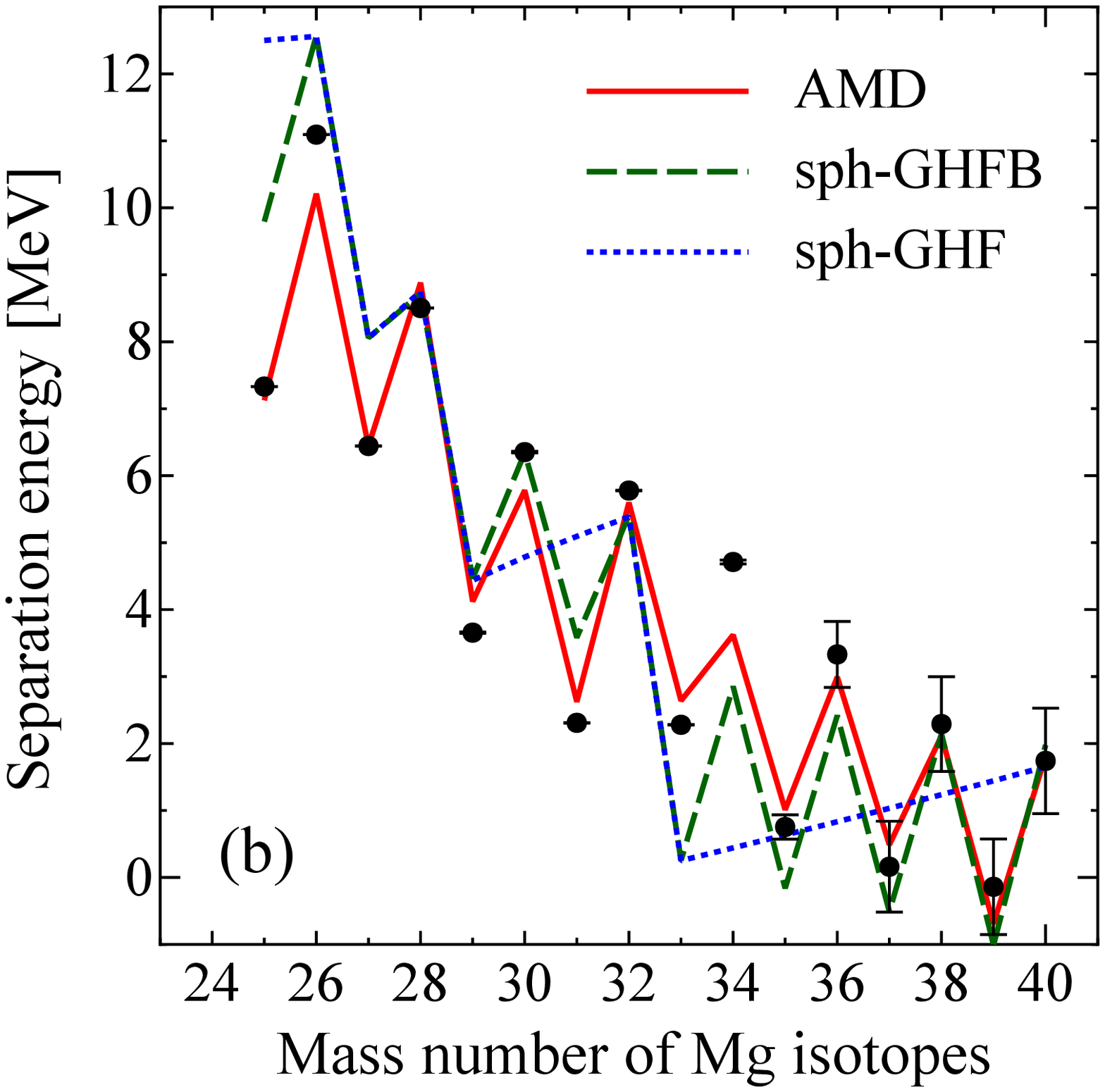}
 \caption{(Color online)
 (a) Total binding energy  per nucleon and (b) one-neutron separation energy
 as a function of mass number for Mg isotopes.
 The solid, dashed and dotted lines represent
 the results of AMD, sph-GHFB and sph-GHF calculations,
 respectively. $^{39}$Mg is unbound in AMD calculations.
 The experimental data are taken from Ref.~\cite{Audi-2012}.
}
 \label{Fig-Sn}
\end{center}
\end{figure}

In Fig.~\ref{Fig-Sn}, total binding energy per nucleon
and $S_{-1n}$ are shown as a function of $A$ for Mg isotopes.
The experimental data~\cite{Audi-2012} are
compared with the results of sph-GHF, sph-GHFB and AMD calculations,
where the Gogny-D1S force is used consistently in all models.
The sph-GHF results (dotted line) underestimate measured
binding energies systematically and do not explain the measured
odd-even staggering of $S_{-1n}$.
These are improved by sph-GHFB calculations
with pairing correlations (dashed line),
though $S_{-1n}$ is negative for $^{35,37,39}$Mg in the calculations.
Meanwhile, AMD calculations (solid line) well reproduce the trend of the experimental data
for both total binding energy and $S_{-1n}$.
The drip-line of Mg is located at $^{40}$Mg and $^{39}$Mg is unbound in 
AMD calculations.
Since the basis wave functions with different configurations around the Fermi
surface are superposed in Eq.~(\ref{eq:AMD-wf}), the dominant effect of the pairing correlation is included in AMD calculations, as confirmed 
from the reasonable reproduction of the even-odd staggering in $S_{-1n}$.
Mg isotopes thus greatly gain the total energies by deformation.

\begin{figure}[htbp]
\begin{center}
 \includegraphics[width=0.40\textwidth,clip]{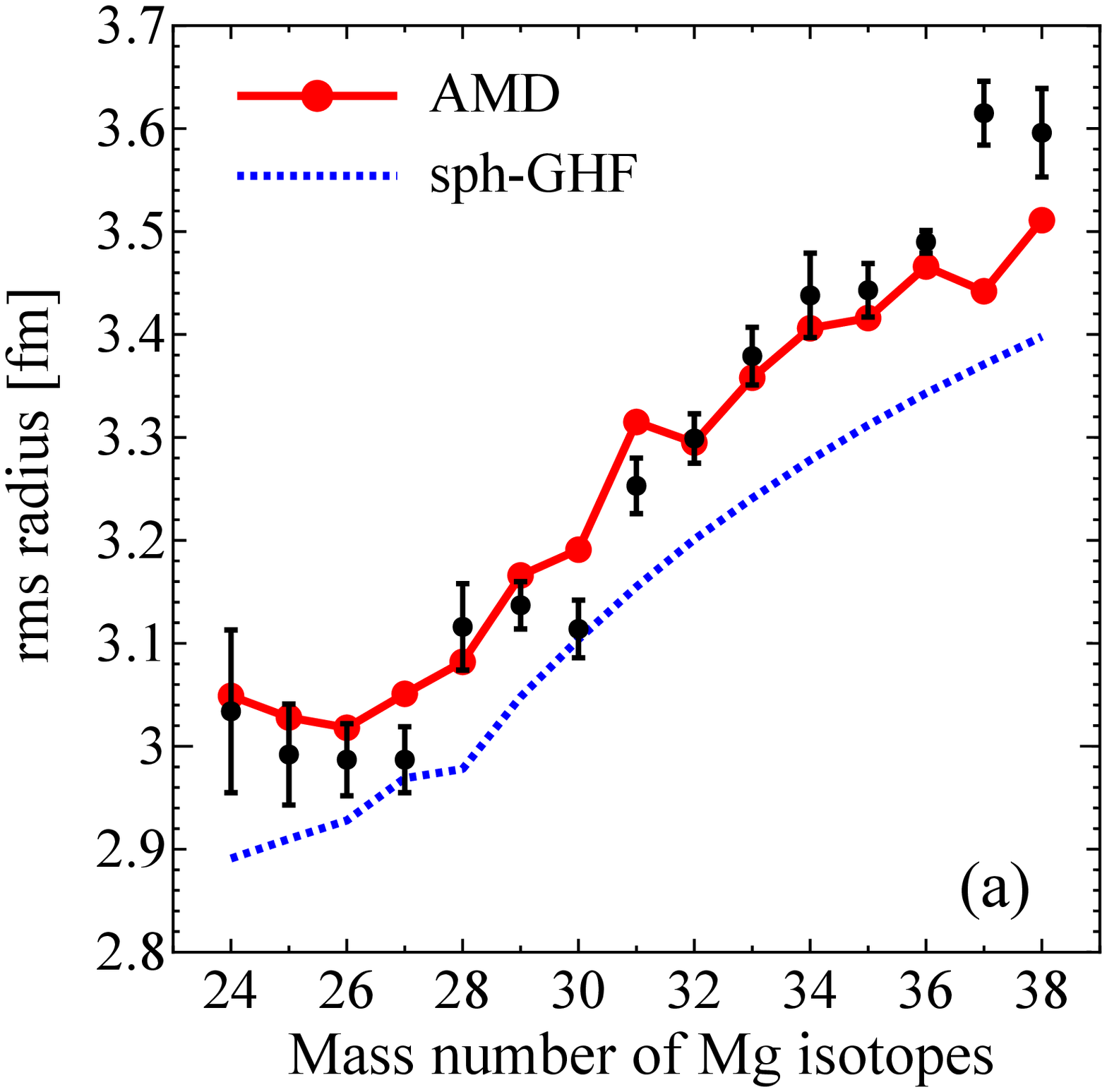}
 \includegraphics[width=0.40\textwidth,clip]{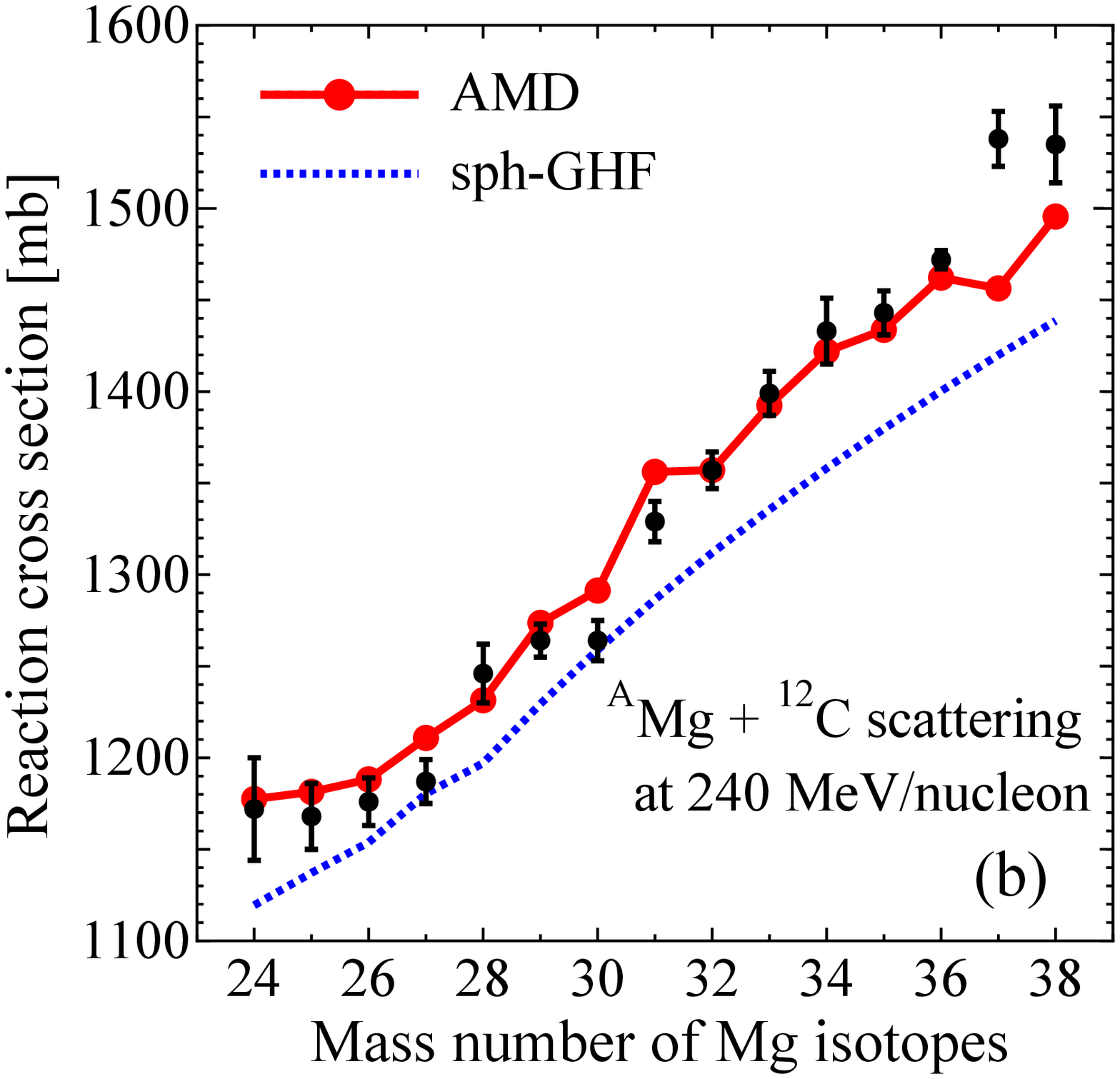}
 \caption{(Color online)
 (a) Matter radii and (b) reaction cross sections $\sigma_{\rm R}$
 as a function of mass number for Mg isotopes.
 The solid and dotted lines represent the results of AMD and sph-GHF
 calculations, respectively.
 The experimental data of $\sigma_{\rm R}$ are taken
from Ref.~\cite{Takechi-Mg}.   }
 \label{Fig-AMD-HF-sigma-R}
\end{center}
\end{figure}

In Fig.~\ref{Fig-AMD-HF-sigma-R}, matter radius and
$\sigma_{\rm R}$ are plotted as a function of $A$
for Mg isotopes. The same discussion can be made between the two quantities.
The AMD results (solid line) yield much better agreement
with the data than the sph-GHF results (dotted line);
note that the effect of pairing correlations is not large
for the matter radius and $\sigma_{\rm R}$, and the sph-GHFB results agree
with the sph-GHF results within the thickness of line.
Deformation enhances matter radius and $\sigma_{\rm R}$
from the dotted to the solid line, indicating the importance of deformation
on matter radius.
The AMD results reproduce the data for $^{24\mbox{--}36}$Mg,
although it considerably
underestimates the data for $^{37,38}$Mg.
Thus AMD predicts ground-state properties of Mg isotopes properly.
Only an exception is the underestimation of matter radius for $^{37,38}$Mg.
This will be discussed in Sec.~\ref{Sec:37Mg}.

\begin{figure}[htbp]
\begin{center}
\includegraphics[width=0.38\textwidth,clip]{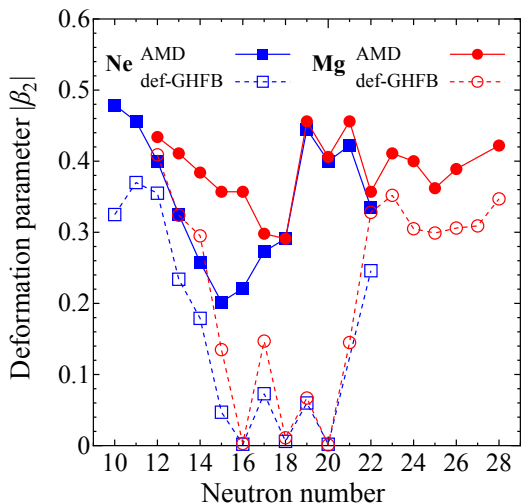}
\caption{(Color online) Theoretical prediction on $|\beta_2|$ for Mg and Ne isotopes.
The solid line with closed circles (squares) are the AMD results with the AMP
for Mg (Ne) isotopes, whereas the dashed line with open circles (squares)
are the deformed Gogny-HFB (def-GHFB) results of Ref.~\cite{HFBwoAMP,HFBchart}
with no AMP for Mg (Ne) isotopes.
}
\label{beta_Ne-Mg}
\end{center}
\end{figure}

In Fig.~\ref{beta_Ne-Mg}, the absolute value of $\beta_2$ is
plotted as a function of neutron number $N$ for Mg and Ne isotopes.
The solid line with closed circles (squares) stands for
the AMD results for Mg (Ne) isotopes;
the values for Ne isotopes are taken from Ref.~\cite{Sumi:2012}.
For both Mg and Ne isotopes, the AMD results show an abrupt increase of
$|\beta_2|$ when $N$ varies from 18 to 19,
where the Nilsson orbitals originating from the spherical $0f_{7/2}$ shell
start to be occupied~\cite{Kimura1}. This indicates that the
island of inversion starts at $N=19$ and the $N=20 $ magicity disappears.
At $N=19 \sim 28$, the $|\beta_2|$ keep
large values of around 0.4. Thus we cannot
identify the endpoint of the island of inversion.
This statement is consistent with the result of in-beam
$\gamma$-ray spectroscopy of $^{34,36,38}$Mg~\cite{Doornenbal-2013}
that the deduced $E(4^+)/E(2^+)$ ratios are about 3.1 independently of $N$.
The $N=28$ magicity, moreover, disappears, since $\beta_2$ is large for
$^{40}$Mg.
The fact that $|\beta_2|$ is large from $^{31}$Mg with $N=19$ to
a drip-line nucleus $^{40}$Mg with $N=28$ indicates that so called
``the island of inversion" may not be an island but a peninsula reaching
the drip line.

Now the AMD results are compared
with the deformed Gogny-HFB (def-GHFB) results of
Refs.~\cite{HFBwoAMP,HFBchart} in Fig.~\ref{beta_Ne-Mg}, where
the dashed line with open circles (squares) denote the def-GHFB results
for Mg (Ne) isotopes.
Note that the AMP is performed in AMD
but not in the def-GHFB calculations of Refs.~\cite{HFBwoAMP,HFBchart}.
Deformation parameter $|\beta_2|$ is enhanced from the def-GHFB results to
the AMD results for both Mg and Ne isotopes. Particularly near and in the
island of inversion, i.e. at $N=16, 18, 20$, the $|\beta_2|$ values are zero
in the def-GHFB results, but they are largely enhanced in the AMD results.
This enhancement is originated from the correlations
induced by the collective rotational motion through the AMP.
These collective-motion effects are particularly
important near and in the island of inversion.

In Ref.~\cite{Horiuchi-2012},
deformed Skyrme Hartree-Fock (def-SHF) calculations were done for Mg isotopes
with two types of interactions SkM$^*$ and SLy4,
where the AMP was not employed.
Around $N=20$, the densities are well deformed with SkM$^*$,
but not with SLy4. The resultant $|\beta_2|$ are then close to
the AMD results when SkM$^*$ interaction is taken,
in spite of the fact that
the AMP is not performed in these def-SHF calculations.

\begin{figure}[htbp]
\begin{center}
\includegraphics[width=0.38\textwidth,clip]{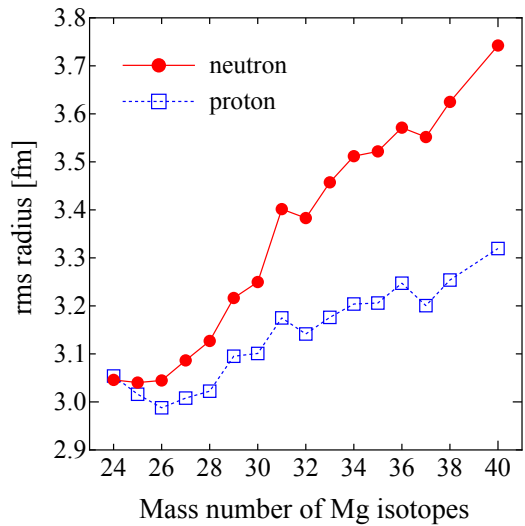}
\caption{(Color online) AMD prediction on neutron and proton rms radii
for Mg isotopes. The solid (dashed) line denotes the neutron (proton) radius.
}
\label{Fig-AMD-HF-RMS_pn}
\end{center}
\end{figure}

\begin{figure}[htbp]
\begin{center}
\includegraphics[width=0.38\textwidth,clip]{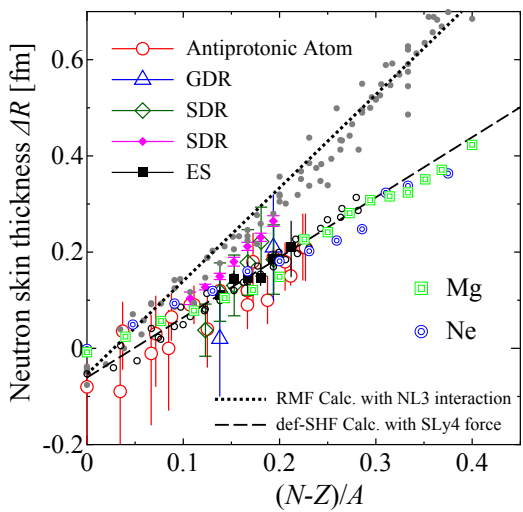}
\caption{(Color online) Neutron skin thickness
$\Delta R$ in Eq.~\eqref{eq:DeltaR}
as a function of asymmetric parameter $(N-Z)/A$.
The double circles and double squares denote the AMD results for
Mg and Ne isotope, respectively. The case of $^{31}$Ne is not plotted here,
since the nucleus has a halo structure.
Neutron skin thickness is also deduced from other measurements
such as antiprotonic atoms (open circles)~\cite{Antiprotnic-Atom},
giant dipole resonance (GDR) (open triangles)~\cite{GDR},
spin dipole resonance (SDR) (open and closed diamonds)~\cite{SDR1,SDR2} and
proton elastic scattering (ES) (closed squares)~\cite{ES1,ES2}.
The AMD results are compared with the results of
relativistic mean-field (RMF) calculations
with NL3 interaction
(small filled dots, whose average trend is represented by the
dotted line)~\cite{RMF},
and of deformed Skyrme-HF (def-SHF) calculations with SLy4 force
(small open dots, whose average trend is represented by the
dashed line)~\cite{Sky-HF}.
}
\label{Fig-nskin}
\end{center}
\end{figure}

Next we compare the neutron rms radius
$\langle \vrr_n^2 \rangle^{1/2}$ with
the proton one ${\langle \vrr_p^2 \rangle}^{1/2}$ in order to
see the isovector component of matter density.
Figure~\ref{Fig-AMD-HF-RMS_pn} shows
$A$-dependence of ${\langle \vrr_n^2 \rangle}^{1/2}$ and
${\langle \vrr_p^2 \rangle}^{1/2}$ for Mg isotopes.
As expected, the neutron skin thickness
\bea
\Delta R\ = {\langle \vrr_n^2 \rangle}^{1/2}
-{\langle \vrr_p^2 \rangle}^{1/2}
\label{eq:DeltaR}
\eea
grows as $A$ increases with $Z$ fixed.
In Fig.~\ref{Fig-nskin}, the $\Delta R$ is
plotted as a function of the asymmetric parameter
$(N-Z)/A$ for Mg and Ne isotopes, where the AMD results
for Ne isotopes are taken from Ref.~\cite{Sumi:2012}.
The present results are consistent with the results of
deformed Skyrme-HF (def-SHF) calculations
with SLy4 interaction (dashed line)~\cite{Sky-HF}
rather than the relativistic mean-field calculations with NL3 interaction
(dotted line)~\cite{RMF}.

\section{Discussions}
\label{Discussions}

\subsection{Comparison between the AMD and def-WS models}
\label{Sec:DWS}

The results of the def-WS model is now compared
with the results of fully-microscopic AMD calculations
in order to see the reliability of the phenomenological model.
The def-WS model yields the same $I^{\pi}$ as AMD for $^{24\mbox{--}40}$Mg except
$^{29}$Mg. For $^{29}$Mg,
the $I^{\pi}$ is $3/2^+$ in AMD, which is consistent with
 the experimental data~\cite{Audi-2012},
 but $1/2^+$ in the def-WS model.
This is because of the difference of the Coriolis coupling in the two models;
its effect is larger in AMD than in def-WS.
In $^{29}$Mg the last odd-neutron occupies the [200 1/2] orbital,
see Table~\ref{tab:DWS-parameter}, which has $\Omega=1/2$
($\Omega$ is the projection of the angular momentum on the symmetry axis).
It is well-known that
the first order Coriolis coupling changes the energy spectrum
of the ${\Omega=1/2}$ rotational band~\cite{BMtextII};
if the effect is strong enough,
the energies of the $I=3/2,7/2,11/2,...$ sequence becomes lower than
those of the $I=1/2,5/2,9/2,...$ sequence.
In the def-WS model the Coriolis coupling can be estimated by
the so-called decoupling parameter; its calculated value is
slightly larger than $-1.0$ in the present case, so that the inversion
between the $I^\pi=3/2^+$ and $I^\pi=1/2^+$ energies does not occur.
If the value is a bit smaller, the $I^\pi=3/2^+$ state
becomes the ground state also in the def-WS model.

\begin{figure}[htbp]
\begin{center}
 \includegraphics[width=0.4\textwidth,clip]{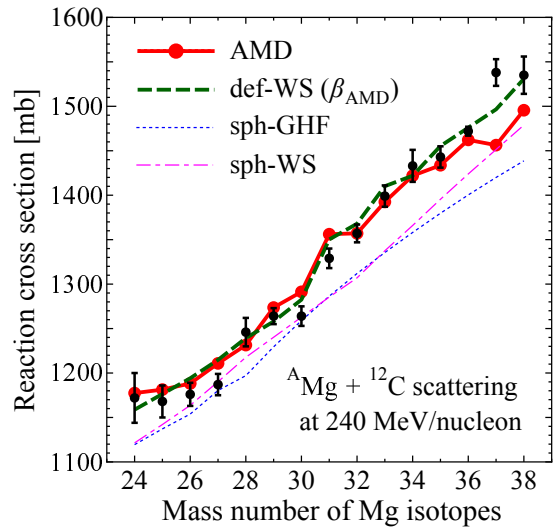}
 \caption{(Color online)
Comparison between AMD and def-WS calculations
in reaction cross sections for Mg isotopes.
The solid (dashed) line denotes the results of AMD (def-WS) calculations,
whereas the thin dotted (dot-dashed) line stands
for the results of sph-GHF (sph-WS) calculations.
The experimental data are taken from Ref.~\cite{Takechi-Mg}.
}
 \label{RCS_Mg_AMD-DWS}
\end{center}
\end{figure}

The results of the AMD and def-WS models are compared for $\sigma_{\rm R}$
in Fig.~\ref{RCS_Mg_AMD-DWS}.
For $^{24\mbox{--}36}$Mg, the def-WS model (dashed line) well simulates
the AMD results (solid line), but  the former overestimates
the latter for $^{37,38}$Mg. To clarify the nature, we also
plot the results of sph-WS and sph-GHF calculations that
correspond to the spherical limit of def-WS and AMD calculations, respectively.
The sph-WS results (thin dot-dashed line) are consistent with
the sph-GHF results (thin dotted line) for $^{24\mbox{--}36}$Mg, but
the former overshoots the latter for $^{37,38}$Mg.
The parametrization of the sph-WS model is thus inappropriate
for $^{37,38}$Mg near the neutron drip line in comparison
with the spherical HF result with Gogny-D1S force.
Some correction should be made in future; in particular,
the depth of potential is assumed to be linear in the asymmetric
parameter $(N-Z)/A$, see Eq.~\eqref{eq:WSdepth}, which may result in
too shallow potentials compared to the sph-GHF for drip line nuclei.
It may be accidental that the def-WS model reproduces 
the measured $\sigma_{\rm R}$ for $^{38}$Mg. 
The fact that AMD calculations underestimate the measured $\sigma_{\rm R}$ 
for $^{37,38}$Mg indicates that these are candidates 
of deformed halo nucleus.

\subsection{Results of AMD-RGM calculation for $^{37}$Mg}
\label{Sec:37Mg}

As shown in Sec.~\ref{Results}, 
the deduced matter radius of $^{37}$Mg is quite large, and at the same time,
the AMD calculation predicts large deformation ($\beta_2 \sim 0.362$)
and very small separation energy ($S_{-1n} = 0.49$ MeV).
These results suggest that $^{37}$Mg is a candidate of deformed halo nucleus.
On the other hand, the matter radius of $^{37}$Mg calculated by AMD
 is much smaller than the one deduced from the measured {$\sigma_{\rm R}$.
This might be because of the inaccuracy of the AMD density in its tail region.}
In this sense, we should solve 
the relative motion between the last neutron and 
the core ($^{36}$Mg) more precisely 
by using the following AMD-RGM framework~\cite{Minomo:2011bb}, 
although the calculations are quite time consuming. 
This procedure is nothing but making a tail correction to AMD density.

In principle, the ground state
$\Phi(^{37}{\rm Mg}; I^\pi)$ of $^{37}$Mg can be expanded in terms of
the ground and excited states $\Phi(^{36}{\rm Mg}; \tilde{I}^{\tilde{\pi}}_i)$
of $^{36}$Mg, where $\tilde{I}^{\tilde{\pi}}_i$ denotes the spin-parity of $^{36}$Mg in its $i$-th state.
This means that 
the ground state of $^{37}$Mg is described by
the $^{36}$Mg + $n$ cluster model with core excitations.
The cluster-model calculation can be done with the RGM 
in which the ground and excited states
of $^{36}$Mg are constructed by AMD:
\begin{align}
 &\Phi(^{37}{\rm Mg}; I^\pi) = \cr
 &\sum_{ilj\tilde{I}\tilde{\pi}}{\cal A}
  \left\{
   R_{ilj}^{}(r) \left[\left[Y_l^{}(\hat{\vrr}) \chi_n^{}\right]_j
    \Phi(^{36}{\rm Mg}; \tilde{I}^{\tilde{\pi}}_i)\right]_{{I}^\pi}
  \right\},\quad
\end{align}
where $\chi_{n}^{}$ is the spin wave function of last neutron and
$R_{ilj}^{}(r)Y_{lm}^{}(\hat{\vrr})$ is the relative wave function
between the last neutron and the core ($^{36}$Mg).
All the excited states of $^{36}$Mg below 8 MeV obtained by the AMD calculation are included as 
$\Phi(^{36}{\rm Mg}; \tilde{I}^{\tilde{\pi}}_i)$.

\begin{figure}[htbp]
\begin{center}
 \includegraphics[width=0.4\textwidth,clip]{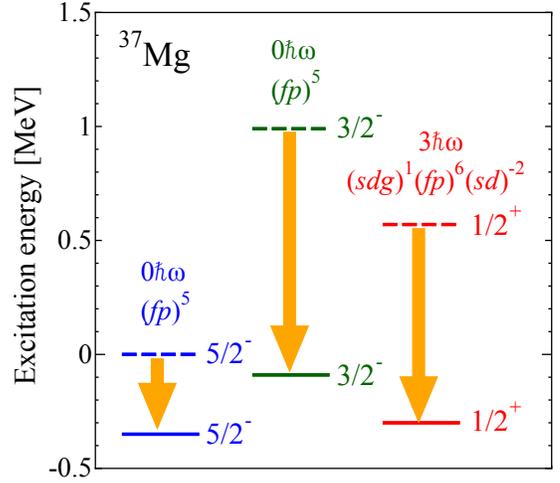}
 \caption{(Color online) Energy spectrum of $^{37}$Mg. 
The dashed (solid) lines denote the results of AMD 
(AMD-RGM) calculations. All the excitation energies
are plotted with reference to the energy of the 5/2- state calculated with AMD.
}
 \label{37Mg-level}
\end{center}
\end{figure}

Figure \ref{37Mg-level} shows energy spectra of $^{37}$Mg calculated 
with the AMD and AMD-RGM models. 
The deviation of the AMD-RGM result (solid line) from the corresponding 
AMD result 
(dashed line) shows an energy gain due to the tail correction. 
Eventually, three bound states appear in the order 
of $I^{\pi}=5/2^{-}$, $1/2^{+}$ and $3/2^{-}$ from the bottom. 
In AMD calculations, the main configuration 
of the $I^{\pi}=1/2^{+}$ state for five valence neutrons 
corresponds to $(sdg)^1(fp)^6(sd)^{-2}$ in the spherical shell model, whereas 
the main configuration of the $I^{\pi}=5/2^{-}$ and $3/2^{-}$ states is 
$(fp)^5$. The $I^{\pi}=1/2^{+}$ state is quite exotic in the sense that 
one neutron is in the $0g1d2s$-shells as a consequence 
of large deformation. These states are quite close in energy, 
and hence there is a possibility that the order is reversed in reality. 
We then assume that anyone of the three states is the ground state.

\begin{figure}[htbp]
\begin{center}
 \includegraphics[width=0.4\textwidth,clip]{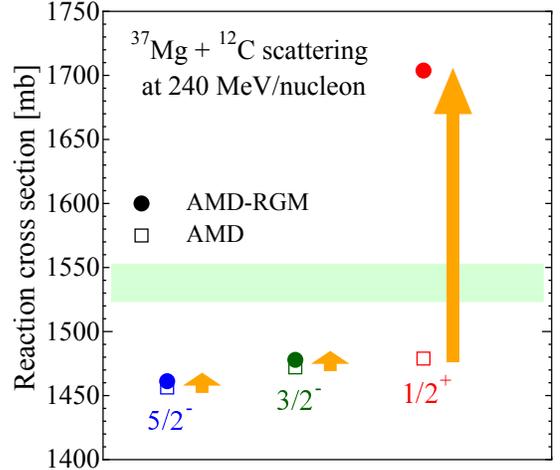}
 \caption{(Color online) Reaction cross sections 
for $^{37}$Mg calculated with the AMD and AMD-RGM methods. 
Open-squares and closed-circles stand for AMD and AMD-RGM results, 
respectively. 
 It is assumed that any one of the $5/2^{-}$, $1/2^{+}$ and $3/2^{-}$ states is 
the ground state.
}
 \label{RCS_AMD-RGM}
\end{center}
\end{figure}

\begin{figure*}[htbp]
\begin{minipage}{0.49\hsize}
\begin{center}
\includegraphics[width=0.8\textwidth,clip]{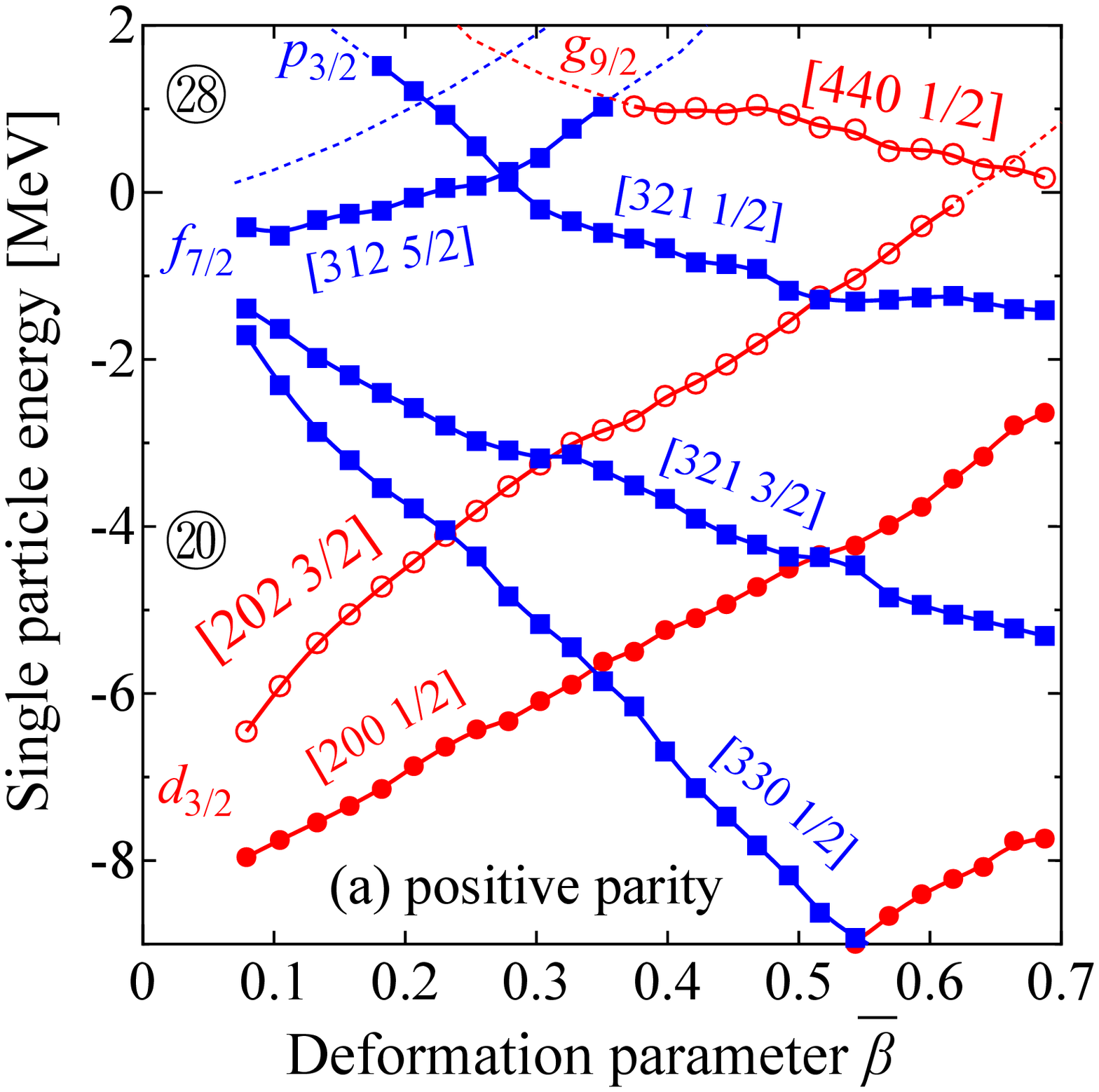}
\end{center}
\end{minipage}
\begin{minipage}{0.49\hsize}
\begin{center}
\includegraphics[width=0.8\textwidth,clip]{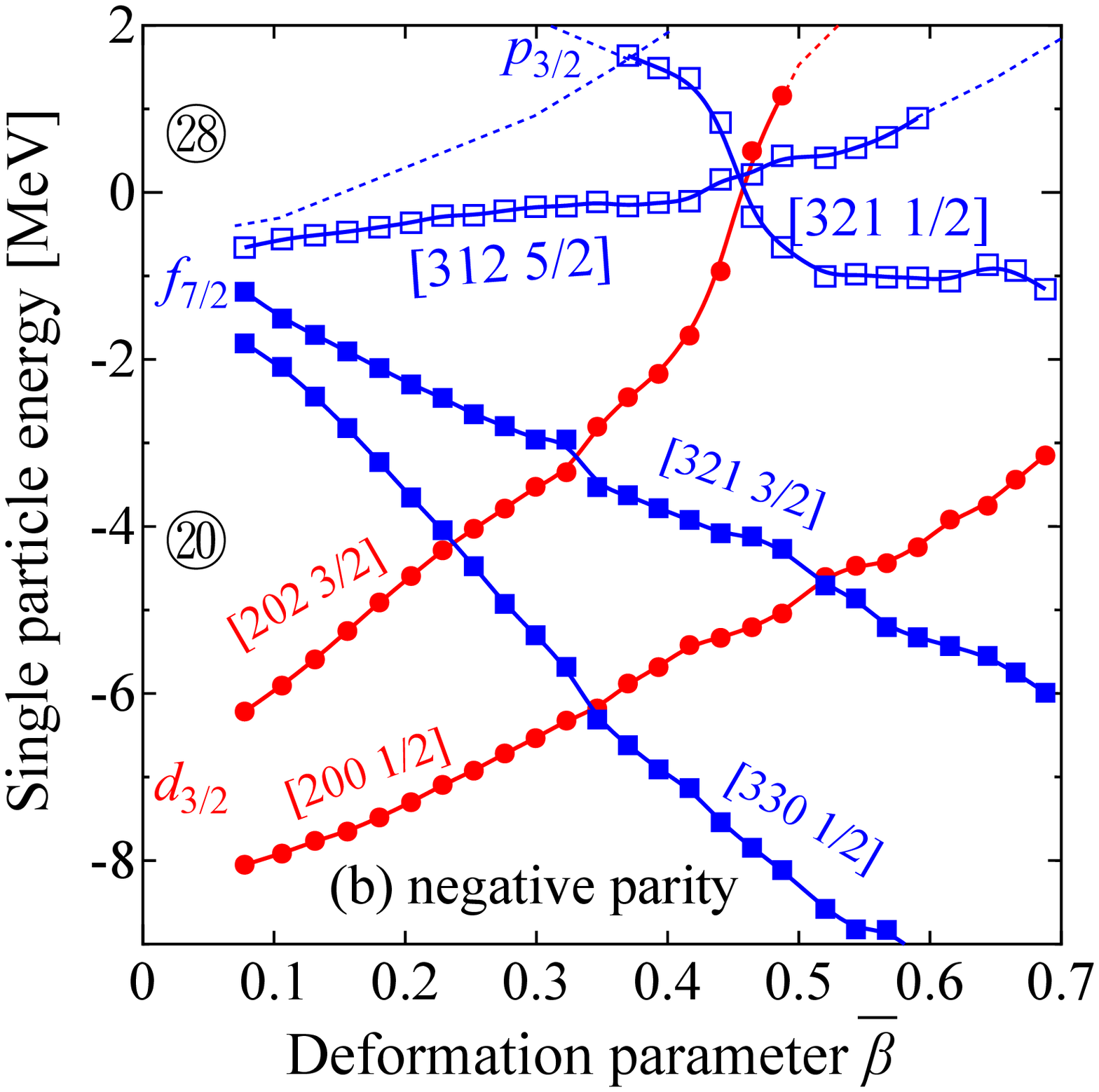}
\end{center}
\end{minipage}
\caption
{Color online) Neutron single-particle energies of $^{37}$Mg
 for (a) positive parity and (b) negative parity.
 Filled (open) symbols mean the orbitals occupied by two (one) neutrons.
Circles (Squares) show the orbital where the amount of the positive-parity component
is larger (smaller) than 50\%.
}
\label{37Mg-Nilsson}
\end{figure*}

In Fig.~\ref{RCS_AMD-RGM}, 
the $\sigma_{\rm R}$ before and after 
the tail correction are shown for the three states.  
When the $5/2^{-}$ state is the ground state, 
the calculated value of $\sigma_{\rm R}$ is little enhanced by the tail 
correction and still 
underestimates the measured $\sigma_{\rm R}$. 
This underestimation may not be resolved even if 
the $5/2^{-}$ state has smaller binding energy, 
since the last neutron is in the [312 5/2] (0$f_{7/2}$-origin) 
orbital; 
see the single particle energy for negative parity 
in Fig.~\ref{37Mg-Nilsson}(b). 
This was numerically confirmed in our previous work~\cite{Takechi-Mg} 
by using the def-WS model, i.e., by changing the parameters of the def-WS 
potential slightly for the last neutron.

When the $1/2^{+}$ state is the ground state, 
the last neutron occupies [440 1/2] (0$g_{9/2}$-origin) as shown 
in Fig.~\ref{37Mg-Nilsson}(a), 
where $s$-wave halo can be formed. 
The calculated value of $\sigma_{\rm R}$ is largely enhanced 
by the tail correction and 
consequently overestimates the measured $\sigma_{\rm R}$. 
However the overestimation can be resolved, 
if the $1/2^{+}$ state has a larger binding energy by some effect. 
When the $3/2^{-}$ state is the ground state, 
the calculated value of $\sigma_{\rm R}$ is little enhanced 
by the tail correction and still 
undershoots the measured $\sigma_{\rm R}$. 
In AMD calculations, the $3/2^{-}$ state has the main component 
in which the last neutron is mainly coupled with not the ground state of 
$^{36}{\rm Mg}$ but the excited $2^{+}$ state, so that 
the last neutron is not weakly bound because of the core excitation. 
Consequently, the $\sigma_{\rm R}$ is hardly enhanced in 
AMD-RGM calculations. 
The $\sigma_{\rm R}$ may be enhanced, if the core excitation 
is suppressed by some effect.

The present AMD-RGM calculations thus cannot reproduce the measured 
$\sigma_{\rm R}$ perfectly. In nuclei near the drip line, in general, 
the kinetic energy is nearly canceled with the potential energy 
coming from the 2-nucleon (2N) central and spin-orbit forces. 
This suggests that higher-order effects 
such as the 2N tensor force and the 3-nucleon (3N) force become 
important. 
In our previous work~\cite{Takechi-Mg} based on the def-WS model, 
the ground-state spin-parity is $I^{\pi}=5/2^{-}$,
since the [312 5/2] (0$f_{7/2}$-origin) orbital is slightly lower in energy 
than the others. To explain the measured large $\sigma_{\rm R}$ for $^{37}$Mg,  we assumed that the last neutron is in 
the [321 1/2] (1$p_{3/2}$-origin) orbital. 
As an underlying mechanism of the inversion,
we can consider the 2N tensor force and the 3N force,
since it is reported that the 2N tensor force reduces
the energy difference between 0$f_{7/2}$ and 1$p_{3/2}$ 
levels in the spherical shell model~\cite{Otsuka-2010}
and the three-body force weakens the strength of the spin-orbit interaction
in neutron-rich nuclei~\cite{Kohno-2012}. 
Further analyses along this line are quite interesting. 
The analysis of $^{38}$Mg is also an important future work after understanding 
the structure of $^{37}$Mg.

\section{Summary}
\label{Summary}

We have determined matter radii of $^{24\mbox{--}38}$Mg
systematically from measured $\sigma_{\rm R}$, fine-tuning the
parameters of the def-WS model.
This flexibility is an advantage of the def-WS model.
The deduced matter radii are largely enhanced from 
the stable and spherical limit estimated by sph-GHF calculations
with the Gogny-D1S interaction for stable spherical nuclei.
Two thirds of the enhancement come from nuclear deformation,
whereas one third is from neutron-skin and/or weak-binding effects.

Fully-microscopic AMD calculations with the Gogny-D1S interaction,
meanwhile, have no free parameter and hence high predictability,
if the calculations are successful in reproducing existing experimental data
systematically.
The AMD calculations well reproduce measured
ground-state properties (spin-parity,
total binding energy and one-neutron separation energy) of Mg isotopes.
The deduced matter radii can be also well reproduced
for $^{24\mbox{--}36}$Mg.
AMD is thus reliable and hence have high predictability.
As for $^{37,38}$Mg,
theoretical matter radii calculated with AMD are enhanced by deformation,
but still considerably underestimate
the deduced matter radii.
This problem is not cured even with the more sophisticated AMD-RGM framework.
Further theoretical investigation should be done; e.g.,
the effective interaction Gogny-D1S force may not be best suited
for the description of drip line nuclei.
This large enhancement of the measured matter radius suggests
that $^{37,38}$Mg are candidates for deformed halo nucleus.

Neutron number ($N$) dependence 
of deformation parameter $\beta_2$ is predicted by AMD.
For both Mg and Ne isotopes, AMD calculations show an abrupt increase
of $\beta_2$ at $N=19$, where the Nilsson orbitals originating
from the spherical $0f_{7/2}$ shell begin to be occupied.
The starting point of the island of inversion is thus $N=19$.
At $N=19 \sim 28$, the $\beta_2$ keep large values of around 0.4.
Hence there seems to be no endpoint of the island of inversion.
Moreover, the $N=20$ and 28 magicities disappear.
Since $^{40}$Mg with $N=28$ may be a drip-line nucleus,
so called ``the island of inversion" may not be an island but a peninsula that
reaches the neutron drip line.
At $N=16, 18$ and 20, deformation parameter $\beta_2$ vanishes
in def-GHFB calculations with no AMP, but becomes large
in AMD calculations with the AMP. The correlations induced by
the collective rotational motion through the AMP are thus important.

Neutron number dependence of neutron skin thickness ($\Delta R$)
is also predicted by AMD. The AMD results for Mg and Ne isotopes
are consistent with the results of Skyrme-HF calculations with SLy4 force
rather than the relativistic mean-field calculations with NL3 interaction.

\vspace*{5mm}

\section*{Acknowledgements}
The authors thank K. Ogata and H. Sakurai for fruitful discussions.
This work is supported in part by Grant-in-Aid for Scientific Research (KAKENHI) from Japan Society for the Promotion of Science
 (25$\cdot$4319, 
 24$\cdot$4137, 
 25$\cdot$949, 
 25400240, 
 22540285), 
 and by Grant-in-Aid for Scientific Research on Innovative Areas from MEXT (2404: 24105008).
The numerical calculations of this work were performed
on the computing system in Research Institute for Information Technology of Kyushu University
and the HITACHI SR16000 at KEK and YITP.

\noindent
\appendix

\section{Woods-Saxon potential parameter set}
\label{Parameter set of Woods-Saxon potential}

In this paper, the strength $V_0$ of the WS potential in Eq.~\eqref{DWS-0}
is parameterized as
\begin{equation}
V_0=-V\times\left(1\pm\kappa\frac{N-Z}{A}\right),
\hspace{3mm}
\left \{
\begin{array}{l}
+\ \mathrm{proton} \\
-\ \mathrm{neutron}
\end{array}
\right.
\label{eq:WSdepth}
\end{equation}
with proton, neutron and mass numbers, $Z$, $N$ and $A$.
The WS potential is then characterized by the parameters,
$V, \kappa, R_0, a$ and $\lambda$.
We use the parameter set provided by Ramon Wyss~\cite{WyssPriv}.
The set is fitted to the moment of inertia and the quadrupole moment systematically for medium and heavy nuclei. The set has already used in some works
with success~\cite{Sumi:2012,Minomo-DWS}.
The parameter set is shown in Table~\ref{tab:DWS-parameter set};
see Ref.~\cite{SS09} for the Coulomb part.

\begin{table*}
\caption{
The parameter set of the Woods-Saxon potential adopted in this work~\cite{WyssPriv}.
As other physical constants, $e^2/(\hbar c)=137.03602$, $\hbar c=197.32891$ MeV$\cdot$fm,
 and $mc^2=938.9059$ MeV are taken.
}
\label{tab:DWS-parameter set}
\begin{center}
\begin{tabular}{cccccccc}\hline \hline
 $V$ [MeV] & $\kappa_\mathrm{c}$ & $\kappa_\mathrm{so}$ & $R_{0\mathrm{c}}$ [fm] & $R_{0\mathrm{so}}$ [fm]        & $a$ [fm] & $\lambda_\mathrm{so}$ \\ \hline
 53.7      & 0.63                & 0.25461              & $1.193A^{1/3}+0.25$    & 0.969$\times R_{0\mathrm{c}}$ & 0.68     & 26.847    \\
 \hline
\end{tabular}
\end{center}
\end{table*}

\section{Relation of deformation parameters between AMD and def-WS models}
\label{Relation of defromation parameters}

We show the relation between $\bar{\beta}$ in AMD and $\beta_2$ in the def-WS
model. For simplicity, we only consider the axially symmetric
deformation ($\gamma=0$).
Equation~\eqref{eq:surf} in Sec.~\ref{Theoretical framework} is reduced to
\begin{align}
 R(\theta) = R_{0}c_{v}(\beta_2)\Bigr[
 1+\beta_2Y_{20}(\theta)\Bigl],
\label{eq:surfA}
\end{align}
where $c_{v}$ is the factor assuring the volume conservation of nucleus.
The relation between $\bar{\beta}$ and $\beta_2$ can be extracted analytically
by considering the sharp-cut density
\begin{equation}
\rho({\bfi r})=\rho_0\theta(R(\theta)-r)
\end{equation}
with $\rho_0=3A/(4\pi R_0^3)$. For this density,
we analytically obtain
\begin{align}
\left<x^2 \right>
&=\int x^2\rho({\bfi r})d{\bfi r}\\
&=\frac{4\rho_0R_0^5c_v^5}{15}
\left(\pi-\frac{\sqrt{5\pi}}{2}\beta_2+\frac{25}{14}\beta_2^2
-\frac{5}{28}\sqrt{\frac{5}{\pi}}\beta_2^3+\cdots\right),
\\
\left<y^2 \right>
&=\left<x^2 \right> ,
\\
\left<z^2 \right>
&=\int z^2\rho({\bfi r})d{\bfi r}
\\
&=\frac{4\rho_0R_0^5c_v^5}{15}
\left(\pi+\sqrt{5\pi}\beta_2+\frac{55}{14}\beta_2^2
+\frac{10}{7}\sqrt{\frac{5}{\pi}}\beta_2^3+\cdots\right).
\end{align}
Combining these equations with Eq.~\eqref{eq:AMDbgz2} leads to
the relation between  $\bar{\beta}$ and $\beta_2$ as
\begin{align}
\bar{\beta}
&=\frac{1}{3}\sqrt{\frac{4\pi}{5}}
\ln\left(\frac{\left<z^2 \right>}{\left<x^2 \right>}\right)\\
&=\beta_2+\frac{1}{28}\sqrt{\frac{5}{\pi}}\beta_2^2
-\frac{25}{28\pi}\beta_2^3+\cdots.
\label{eq:beta_tran}
\end{align}
The polynomial expression up to third order works well for
$-0.6\leqq\beta_2\leqq0.6$.
Obviously, we get $\bar{\beta}=\beta_2$ for
small deformation.

Table~\ref{tab:DWS-parameter} lists up the AMD deformation parameter set
$(\bar{\beta},\bar{\gamma})$ and the corresponding standard set
$(\beta_2,\gamma)$ in the def-WS model for Mg isotopes.
For $^{24\mbox{--}38}$Mg, only $^{30}$Mg has nonzero $\bar{\gamma}$, but
we simply set $\gamma$ to zero.
This procedure is justified, since $\gamma$ deformation little
affects $\sigma_{\rm R}$~\cite{Sumi:2012}.

\begin{table}
\caption{
The deformation parameter set $(\bar{\beta},\bar{\gamma})$ in AMD and
the corresponding standard set $(\beta_2,\gamma)$ in the def-WS model
for Mg isotopes.
The Nilsson asymptotic quantum numbers of last neutron are listed up
in the last column.}
\label{tab:DWS-parameter}
\begin{center}
\begin{threeparttable}
\begin{tabular}{cccccc}\hline \hline
 nuclide   & $\bar{\beta}$ & $\bar{\gamma}$ &  $\beta_2$ & $\gamma$
 & [$N$,$n_3$,$\Lambda$,$\Omega$] for last-$n$ \\ \hline
 $^{24}$Mg & 0.42  &  0$^{^\circ}$  & 0.434 & 0 $^{^\circ}$ & [211 3/2] \\
 $^{25}$Mg & 0.40  &  0$^{^\circ}$  & 0.411 & 0 $^{^\circ}$ & [202 5/2] \\
 $^{26}$Mg & 0.375 &  0$^{^\circ}$  & 0.384 & 0 $^{^\circ}$ & [202 5/2] \\
 $^{27}$Mg & 0.35  &  0$^{^\circ}$  & 0.357 & 0 $^{^\circ}$ & [211 1/2] \\
 $^{28}$Mg & 0.35  &  0$^{^\circ}$  & 0.357 & 0 $^{^\circ}$ & [211 1/2] \\
 $^{29}$Mg & 0.295 &  0$^{^\circ}$  & 0.298 & 0 $^{^\circ}$ & [200 1/2] \\
 $^{30}$Mg & 0.285 & 25$^{^\circ}$  & \hspace{2mm}0.291\tnote{a}\hspace{2mm} & 25.8 $^{^\circ}$ & [200 1/2] \\
 $^{31}$Mg & 0.44  &  0$^{^\circ}$  & 0.456 & 0 $^{^\circ}$ & [200 1/2] \\
 $^{32}$Mg & 0.395 &  0$^{^\circ}$  & 0.406 & 0 $^{^\circ}$ & [200 1/2] \\
 $^{33}$Mg & 0.44  &  0$^{^\circ}$  & 0.456 & 0 $^{^\circ}$ & [321 3/2] \\
 $^{34}$Mg & 0.35  &  0$^{^\circ}$  & 0.357 & 0 $^{^\circ}$ & [321 3/2] \\
 $^{35}$Mg & 0.40  &  0$^{^\circ}$  & 0.411 & 0 $^{^\circ}$ & [202 3/2] \\
 $^{36}$Mg & 0.39  &  0$^{^\circ}$  & 0.400 & 0 $^{^\circ}$ & [202 3/2] \\
 $^{37}$Mg & 0.355 &  0$^{^\circ}$  & 0.362 & 0 $^{^\circ}$ & [312 5/2] \\
 $^{38}$Mg & 0.38  &  0$^{^\circ}$  & 0.389 & 0 $^{^\circ}$ & [312 5/2] \\
 \hline
\end{tabular}
\begin{tablenotes}\footnotesize
\item[a] This value is obtained by assuming $\bar{\gamma}$ is finite.
The value will become 0.288 if $\bar{\gamma}=0$; see Ref. \cite{Sumi:2012} for the detail.
\end{tablenotes}
\end{threeparttable}
\end{center}
\end{table}



\begin{thebibliography}{00}
\bibitem{Klapisch-1969}
R. Klapisch {\it et al.},
Phys. Rev. Lett. {\bf 23}, 652 (1969).

\bibitem{Thibault-1975}
C. Thibault {\it et al.},
Phys. Rev. C {\bf 12}, 644 (1975).

\bibitem{Warburton}
  E.~K.~Warburton, J.~A.~Becker, and B.~A.~Brown,
    Phys.\ Rev.\ C {\bf 41}, 1147 (1990).

\bibitem{Orr-1991}
N. Orr {\it et al.},
Phys. Lett. B {\bf 258}, 29 (1991).

\bibitem{Mot95}
T. Motobayashi {\it et al.},
Phys.\ Lett.\ B {\bf 346}, 9 (1995).

\bibitem{Caurier}
  E.~Caurier, F. Nowacki, A. Poves, and J. Retamosa,
  Phys.\ Rev.\ C {\bf 58}, 2033 (1998).

\bibitem{Utsuno}
  Y.~Utsuno, T. Otsuka, T. Mizusaki, and M. Honma,
  Phys.\ Rev.\ C {\bf 60}, 054315 (1999).

\bibitem{Iwas01}
H. Iwasaki {\it et al.},
Phys.\ Lett.\ B {\bf 522}, 227 (2001).

\bibitem{Yana03}
Y. Yanagisawa {\it et al.},
Phys.\ Lett.\ B {\bf 566}, 84 (2003).

\bibitem{Otsuka-2005}
T. Otsuka, T. Suzuki, R. Fujimoto, H. Grawe, and Y. Akaishi,
Phys. Rev. Lett. {\bf 95}, 232502 (2005).

\bibitem{Otsuka-2010}
T. Otsuka, T. Suzuki, M. Honma, Y. Utsuno, N. Tsunoda, K. Tsukiyama, and M. Hjorth-Jensen,
Phys.\ Lett.\ {\bf 104}, 104, 012501 (2010).


\bibitem{Tanihata}
  I.~Tanihata \textit{et al}.,
  \newblock
  Phys.\ Lett.\ B {\bf 289}, 261 (1992). \\
  I.~Tanihata, \newblock
   J.\ Phys.\ G {\bf 22}, 157 (1996).

\bibitem{Jensen}
  A.~S.~Jensen \textit{et al}., \newblock
    Rev.\ Mod.\ Phys.\ {\bf 76}, 215 (2004).

\bibitem{Jonson}
   B. Jonson, \newblock
   Phys.\ Rep.\ {\bf 389}, 1 (2004).

\bibitem{Sumi:2012}
  T.~Sumi, K.~Minomo, S.~Tagami, M.~Kimura, T. Matsumoto, K.~Ogata, Y.~R.~Shimizu, and M.~Yahiro,
  Phys. Rev. C {\bf 85}, 064613 (2012).

\bibitem{Nakamura}
    T.~Nakamura \textit{et al}.,
    Phys. Rev. Lett. {\bf 103}, 262501 (2009).

\bibitem{Takechi}
M. Takechi {\it et al.}, Phys. Lett. B {\bf 707}, 357 (2010).

\bibitem{Takechi-Mg}
M. Takechi {\it et al.}, submitted to Phys. Rev. Lett.,
EPJ Web of Conferences {\bf 66}, 02101 (2014).

\bibitem{M3Y}
G. Bertsch, J. Borysowicz, H. McManus, and W.G. Love,
Nucl. Phys. A{\bf 284}, 399 (1977).

\bibitem{JLM}
J.-P. Jeukenne, A. Lejeune and C. Mahaux, Phys. Rev. C{\bf 16}, 80 (1977);
ibid. Phys. Rep. {\bf 25}, 83 (1976).


\bibitem{Brieva-Rook}
F.A. Brieva and J.R. Rook, Nucl. Phys. A{\bf 291}, 299 (1977);
ibid. 291, 317 (1977); ibid. 297, 206 (1978).

\bibitem{Satchler-1979}
G. R. Satchler and W. G. Love, Phys. Rep. {\bf 55}, 183-254 (1979).

\bibitem{Satchler}
G. R. Satchler, "Direct Nuclear Reactions",
Oxfrod University Press, (1983).

\bibitem{CEG}
N. Yamaguchi, S. Nagata, and T. Matsuda, Prog. Theor.
Phys. {\bf 70}, 459 (1983);
N. Yamaguchi, S. Nagata, and J. Michiyama,
Prog. Theor. Phys. {\bf 76}, 1289 (1986).

\bibitem{Rikus-von Geramb}
L. Rikus, K. Nakano, and H. V. von Geramb, Nucl. Phys. A{\bf 414}, 413 (1984);
L. Rikus and H.V. von Geramb, Nucl. Phys. A{\bf 426}, 496 (1984).

\bibitem{Amos}
K. Amos, P. J. Dortmans, H. V. von Geramb, S. Karataglidis,
and J. Raynal, in \textit{Advances in Nuclear Physics}, edited by
J. W. Negele and E. Vogt(Plenum, New York, 2000) Vol. 25, p. 275.

\bibitem{rainbow}
 D. T. Khoa,
 W. von Oertzen, H. G. Bohlen, and S. Ohkubo,
 J.\ Phys.\ G {\bf 34}, R111-R164 (2007).

\bibitem{CEG07}
T. Furumoto, Y. Sakuragi, and Y. Yamamoto, Phys. Rev. C{\bf 78},
044610 (2008); {\it ibid.}, C{\bf 79}, 011601(R) (2009);
{\it ibid.}, C{\bf 80}, 044614 (2009).

\bibitem{Toyokawa:2013uua}
  M.~Toyokawa, K.~Minomo, and M.~Yahiro,
 Phys. Rev. C{\bf 88}, 054602 (2013),

\bibitem{ERT}
M. Yahiro, K. Ogata, and K. Minomo,
Prog. Theor. Phys. {\bf 126}, 167 (2011).

\bibitem{GognyD1S}
J.~F.~Berger, M.~Girod, and D.~Gogny,
Comput. Phys. Commun. {\bf 63}, 365 (1991).


\bibitem{Kimura}
M. Kimura and H. Horiuchi, Prog. Theor. Phys. 111, 841 (2004).

\bibitem{Kimura1}
M. Kimura, Phys. Rev. C{\bf 75}, 041302 (2007).

\bibitem{RER02}
R.~Rodr\'iguez-Guzm\'an, J.L. Egido, and L.M. Robledo,
Nucl.\ Phys.\ A {\bf 709}, 201 (2002).

\bibitem{RER03}
R.R.~Rodr\'iguez-Guzm\'an, J.L. Egido, and L.M. Robledo,
Eur.\ Phys.\ J.\ A {\bf 17}, 37 (2003).

\bibitem{Minomo:2011bb}
  K.~Minomo, T.~Sumi, M.~Kimura, K.~Ogata, Y.~R.~Shimizu, and M.~Yahiro,
  Phys. Rev. Lett. {\bf 108}, 052503 (2012).

\bibitem{Minomo-DWS}
K. Minomo, T. Sumi, M. Kimura, K. Ogata, Y. R. Shimizu, and M. Yahiro,
Phys. Rev. C {\bf 84}, 034602 (2011).



\bibitem{Watson}
K. M. Watson, Phys. Rev. {\bf 89}, 575 (1953).

\bibitem{KMT}
A. K. Kerman, H. McManus, and R. M. Thaler, Ann.
Phys. {\bf 8}, 551 (1959).

\bibitem{Yahiro-Glauber}
M.~Yahiro, K.~Minomo, K.~Ogata, and M.~Kawai,
Prog.\ Theor.\ Phys.\ {\bf 120}, 767 (2008).

\bibitem{DFM-standard-form}
B. Sinha, Phys. Rep. {\bf 20}, 1 (1975). \\
B. Sinha and S. A. Moszkowski, Phys. Lett. B{\bf 81}, 289 (1979).

\bibitem{DFM-standard-form-2}
T. Furumoto, Y. Sakuragi, and Y. Yamamoto, Phys. Rev. C{\bf 82}, 044612 (2010).

\bibitem{Minomo:2009ds}
  K.~Minomo, K.~Ogata, M.~Kohno, Y.~R.~Shimizu, and M.~Yahiro,
  J.\ Phys.\ G {\bf 37}, 085011 (2010)
  [arXiv:0911.1184 [nucl-th]].

\bibitem{Hag06}
K. Hagino, T. Takehi, and N. Takigawa,
Phys. Rev. C {\bf 74} (2006), 037601.



%

\bibitem{WyssPriv}
R. Wyss, private communication (2005).

\bibitem{Nakada:2008}
H. Nakada,
Nucl. Phys. \textbf{A808} 47, (2008).





\bibitem{C12-density}
H. de Vries, C. W. de Jager, and C. de Vries,
At. Data Nucl. Data Tables \textbf{36}, 495 (1987).

\bibitem{expC12C12}
 M. Takechi \textit{et al}.,
  Phys.\ Rev.\ C \textbf{79}, 061601(R) (2009).

\bibitem{Ne20-sigmaI}
L. Chulkov \textit{et al}.,
Nucl. Phys. \textbf{A603} 219, (1996).

\bibitem{Na23-sigmaI}
T. Suzuki \textit{et al}.,
Phys. Rev. Lett. \textbf{75}, 3241 (1995).

\bibitem{Kox-Ca-1984}
S. Kox {\it et al.}, Nuclear Physics A{\bf 420}, 162 (1984).

\bibitem{Audi-2012}
G. Audi \textit{et al}.,
Chinese Physics C {\bf 36}, 1157 (2012).

\bibitem{Baumann-2007}
T. Baumann {\it et al.}, Nature {\bf 449}, 1022 (2007).

\bibitem{Doornenbal-2013}
P. Doornenbal \textit{et al}.,
Phys. Rev. Lett. \textbf{111}, 212502 (2013).

\bibitem{HFBwoAMP}
S. Hilaire and M. Girod, Eur. Phys. J. A \textbf{33}, 237 (2007).

\bibitem{HFBchart}
http://www-phynu.cea.fr/science\_en\_ligne/
\\
carte\_potentiels\_microscopiques/
\\
carte\_potentiel\_nucleaire\_eng.htm

\bibitem{Horiuchi-2012}
W. Horiuchi, T. Inakura, T. Nakatsukasa, and Y. Suzuki,
Phys. Rev. C{\bf 86}, 024614 (2012).

\bibitem{Antiprotnic-Atom}
A. Trzci$\acute{\mathrm{n}}$ska \textit{et al}.,
Phys. Rev. Lett. \textbf{87}, 082501 (2001).

\bibitem{GDR}
A. Krasznahorkay \textit{et al}.,
Nucl. Phys. \textbf{A567}, 521, (1994).

\bibitem{SDR1}
A. Krasznahorkay \textit{et al}.,
Phys. Rev. Lett. \textbf{82}, 3216 (1999).

\bibitem{SDR2}
A. Krasznahorkay \textit{et al}.,
Nucl. Phys. \textbf{A731}, 224, (2004).

\bibitem{ES1}
S. Terashima \textit{et al}.,
Phys. Rev. C{\bf 77}, 024317 (2008).

\bibitem{ES2}
J. Zenihiro \textit{et al}.,
Phys. Rev. C{\bf 82}, 044611 (2010).

\bibitem{Sky-HF}
P. Sarriguren, M. K. Gaidarov, E. Moya de Guerra, and A. N. Antonov,
Phys. Rev. C{\bf 76}, 044322 (2007).

\bibitem{RMF}
G.A. Lalazissis, D. Vretenar, and P. Ring,
Phys. Rev. C{\bf 57}, 2294 (1998).



\bibitem{BMtextII}
A.~Bohr and B.~R.~Mottelson,
{\it Nuclear Structure}, Vol.~II Benjamin, New York (1975).

\bibitem{Kohno-2012}
M. Kohno, Phys.\ Rev.\ C {\bf 86}, 061301 (2012).


\bibitem{SS09}
T. Shoji and Y. R. Shimizu,
Prog.\ Theor.\ Phys.\ {\bf 121}, 319 (2009).

\end{thebibliography}
\end{document}